\documentclass[twocolumn,showpacs,aps,prc,floatfix,superscriptaddress]{revtex4}
\usepackage{graphicx}
\usepackage{psfig}
\usepackage{color}

\newcommand{ \be }{\begin{equation}}
\newcommand{ \ee }{\end{equation}}
\newcommand{ \bea }{\begin{eqnarray}}
\newcommand{ \eea }{\end{eqnarray}}

\begin{document}
\title{
Measurements of $\phi$ meson production in relativistic heavy-ion
collisions at RHIC }
\affiliation{Argonne National Laboratory, Argonne, Illinois 60439, USA}
\affiliation{University of Birmingham, Birmingham, United Kingdom}
\affiliation{Brookhaven National Laboratory, Upton, New York 11973, USA}
\affiliation{University of California, Berkeley, California 94720, USA}
\affiliation{University of California, Davis, California 95616, USA}
\affiliation{University of California, Los Angeles, California 90095, USA}
\affiliation{Universidade Estadual de Campinas, Sao Paulo, Brazil}
\affiliation{Carnegie Mellon University, Pittsburgh, Pennsylvania 15213, USA}
\affiliation{University of Illinois at Chicago, Chicago, Illinois 60607, USA}
\affiliation{Creighton University, Omaha, Nebraska 68178, USA}
\affiliation{Nuclear Physics Institute AS CR, 250 68 \v{R}e\v{z}/Prague, Czech Republic}
\affiliation{Laboratory for High Energy (JINR), Dubna, Russia}
\affiliation{Particle Physics Laboratory (JINR), Dubna, Russia}
\affiliation{Institute of Physics, Bhubaneswar 751005, India}
\affiliation{Indian Institute of Technology, Mumbai, India}
\affiliation{Indiana University, Bloomington, Indiana 47408, USA}
\affiliation{Institut de Recherches Subatomiques, Strasbourg, France}
\affiliation{University of Jammu, Jammu 180001, India}
\affiliation{Kent State University, Kent, Ohio 44242, USA}
\affiliation{University of Kentucky, Lexington, Kentucky, 40506-0055, USA}
\affiliation{Institute of Modern Physics, Lanzhou, China}
\affiliation{Lawrence Berkeley National Laboratory, Berkeley, California 94720, USA}
\affiliation{Massachusetts Institute of Technology, Cambridge, MA 02139-4307, USA}
\affiliation{Max-Planck-Institut f\"ur Physik, Munich, Germany}
\affiliation{Michigan State University, East Lansing, Michigan 48824, USA}
\affiliation{Moscow Engineering Physics Institute, Moscow Russia}
\affiliation{City College of New York, New York City, New York 10031, USA}
\affiliation{NIKHEF and Utrecht University, Amsterdam, The Netherlands}
\affiliation{Ohio State University, Columbus, Ohio 43210, USA}
\affiliation{Panjab University, Chandigarh 160014, India}
\affiliation{Pennsylvania State University, University Park, Pennsylvania 16802, USA}
\affiliation{Institute of High Energy Physics, Protvino, Russia}
\affiliation{Purdue University, West Lafayette, Indiana 47907, USA}
\affiliation{Pusan National University, Pusan, Republic of Korea}
\affiliation{University of Rajasthan, Jaipur 302004, India}
\affiliation{Rice University, Houston, Texas 77251, USA}
\affiliation{Universidade de Sao Paulo, Sao Paulo, Brazil}
\affiliation{University of Science \& Technology of China, Hefei 230026, China}
\affiliation{Shanghai Institute of Applied Physics, Shanghai 201800, China}
\affiliation{SUBATECH, Nantes, France}
\affiliation{Texas A\&M University, College Station, Texas 77843, USA}
\affiliation{University of Texas, Austin, Texas 78712, USA}
\affiliation{Tsinghua University, Beijing 100084, China}
\affiliation{United States Naval Academy, Annapolis, MD 21402, USA}
\affiliation{Valparaiso University, Valparaiso, Indiana 46383, USA}
\affiliation{Variable Energy Cyclotron Centre, Kolkata 700064, India}
\affiliation{Warsaw University of Technology, Warsaw, Poland}
\affiliation{University of Washington, Seattle, Washington 98195, USA}
\affiliation{Wayne State University, Detroit, Michigan 48201, USA}
\affiliation{Institute of Particle Physics, CCNU (HZNU), Wuhan 430079, China}
\affiliation{Yale University, New Haven, Connecticut 06520, USA}
\affiliation{University of Zagreb, Zagreb, HR-10002, Croatia}

\author{B.~I.~Abelev}\affiliation{University of Illinois at Chicago, Chicago, Illinois 60607, USA}
\author{M.~M.~Aggarwal}\affiliation{Panjab University, Chandigarh 160014, India}
\author{Z.~Ahammed}\affiliation{Variable Energy Cyclotron Centre, Kolkata 700064, India}
\author{B.~D.~Anderson}\affiliation{Kent State University, Kent, Ohio 44242, USA}
\author{D.~Arkhipkin}\affiliation{Particle Physics Laboratory (JINR), Dubna, Russia}
\author{G.~S.~Averichev}\affiliation{Laboratory for High Energy (JINR), Dubna, Russia}
\author{Y.~Bai}\affiliation{NIKHEF and Utrecht University, Amsterdam, The Netherlands}
\author{J.~Balewski}\affiliation{Massachusetts Institute of Technology, Cambridge, MA 02139-4307, USA}
\author{O.~Barannikova}\affiliation{University of Illinois at Chicago, Chicago, Illinois 60607, USA}
\author{L.~S.~Barnby}\affiliation{University of Birmingham, Birmingham, United Kingdom}
% ...

\collaboration{STAR Collaboration}\noaffiliation
\begin{abstract}
We present results for the measurement of $\phi$ meson production
via its charged kaon decay channel $\phi \rightarrow K^+K^-$ in
Au+Au collisions at $\sqrt{s_{_{NN}}}=62.4$, 130, and 200~GeV, and in
$p+p$ and $d$+Au collisions at $\sqrt{s_{_{NN}}}=200$~GeV from the STAR
experiment at the BNL Relativistic Heavy Ion Collider (RHIC).  The midrapidity ($|y|<0.5$) $\phi$ meson transverse momentum ($p_{T}$) spectra in central Au+Au collisions are found to be well described by a single exponential
distribution. On the other hand, the $p_{T}$ spectra from $p+p$, $d$+Au and peripheral Au+Au collisions show power-law tails at intermediate and high $p_{T}$ and are described better by Levy distributions.
The constant $\phi/K^-$ yield ratio vs beam
species, collision centrality and colliding energy is in
contradiction with expectations from models having kaon coalescence
as the dominant mechanism for $\phi$ production at RHIC. The
$\Omega/\phi$ yield ratio as a function of $p_{T}$ is consistent
with a model based on the recombination of thermal $s$ quarks up to
$p_{T}\sim 4$ GeV/$c$, but disagrees at higher transverse momenta.
The measured nuclear modification factor, $R_{dAu}$, for the $\phi$ meson 
increases above unity at intermediate $p_{T}$, similar to that for pions and protons, 
while $R_{AA}$ is suppressed due to the energy loss effect in central Au+Au collisions. 
Number of constituent quark scaling of both $R_{cp}$ and
$v_{2}$ for the $\phi$ meson with respect to other hadrons in
Au+Au collisions at $\sqrt{s_{_{NN}}}$=200~GeV at intermediate
$p_{T}$ is observed. These observations support quark coalescence as
being the dominant mechanism of hadronization in the intermediate
$p_{T}$ region at RHIC.
\end{abstract}

\maketitle

\begin{center}
\textbf{I.~INTRODUCTION}
\end{center}

The $\phi$(1020) vector meson's properties and its transport in the
nuclear medium have been of interest since its
discovery~\cite{Bertanza:1962}. The proper lifetime of the $\phi$
meson is about 45 fm/c and it decays into charged kaons $K^{+}
K^{-}$ with a branching ratio of $49.2 \%$, and more rarely into
the dilepton pairs $e^{+} e^{-}$ (B. R. of $2.97\times
10^{-4}$) and $\mu^{+} \mu^{-}$ (B. R. of $2.86\times 10^{-4}$).

The mechanism for $\phi$ meson production in high energy collisions
has remained an open issue. As the lightest bound state of strange
quarks ($s\bar{s}$) with hidden strangeness, $\phi$ meson production
is suppressed in elementary collisions because of the
Okubo-Zweig-Iizuka (OZI) rule~\cite{Okubo:1963, Zweig:1964,
Iizuka:1966}. The OZI rule states that processes with disconnected
quark lines in the initial and final state are
suppressed. In an
environment with many strange quarks, $\phi$ mesons can be produced
readily through coalescence, bypassing the OZI
rule~\cite{Shor:1984ui}. The $\phi$ meson has
been predicted to be a probe of the quark-gluon plasma (QGP) formed in ultrarelativistic
heavy-ion collisions \cite{Rafelski:1982pu,
Rafelski:1983hg,Jacob:1982qk, Koch:1984tz,Baltz:1995tv,
Sorge:1992ej}.

On the other hand, a naive interpretation of
$\phi$ meson enhancement in heavy-ion collisions would be that the $\phi$ meson is produced via $K \bar{K} \rightarrow \phi$ in the hadronic
rescattering stage. Models that include hadronic rescatterings
such as RQMD~\cite{Sorge:1995dp} and UrQMD~\cite{Bleicher:1999xi}
have predicted an increase of the $\phi$ to $K^-$
production ratio at midrapidity as a function of the
number of participant nucleons. This prediction was disproved in Au+Au collisions at
$\sqrt{s_{_{NN}}}=200$~GeV, from STAR year 2001 data~\cite{Adams:2004ux}.
With the higher statistics data newly recorded from the Solenoidal Tracker at RHIC (STAR) experiment~\cite{Ackermann:2002ad},
a precise measurement of the $\phi/K^-$ ratio as a function of beam energy and collision centrality is presented to confirm this finding in this paper. 

The in-medium properties of vector mesons in the hot and dense environment are also interesting~\cite{Sorge:1992ej}. The mass and width of the $\phi$ meson were predicted to change because of the partial restoration of chiral symmetry in the nuclear medium. Asakawa and Ko~\cite{Asakawa:1994tp} and Song~\cite{Song:1996gw} predicted that the $\phi$ mass
decreases as a result of many-body effects in a hadronic medium. A double $\phi$ peak
structure in the dilepton invariant mass spectrum from relativistic heavy-ion collisions was proposed as a
signature of a phase transition from the QGP to hadronic matter~\cite{Asakawa:1994nn}. Other calculations have
predicted that the $\phi$ meson width can be widened significantly due to nuclear medium
effects~\cite{Haglin:1994xu, Smith:1997xu, Ko:1993id}. Recently, an interesting $\phi$ mass modification at
normal nuclear density in 12~GeV $p+A$ interactions was observed in the dilepton channel ($e^{+}+e^{-}$) from
the KEK experiment~\cite{Muto:2005za, sakuma:152302}. In STAR, we measure the $\phi$ meson mass and width to compare with these
predictions/measurements, using the decay channel $\phi$$\rightarrow$$K^{+}K^{-}$.

From phenomenological analysis, it is suggested that the $\phi$
meson mean free path in hadronic media is large because of its small
cross section of scattering with hadrons~\cite{Shor:1984ui}. Many other calculations also indicate
that the $\phi$ meson has small rescattering cross sections
with hadronic matter~\cite{Haglin:1994xu, Smith:1997xu}. However,
after including three- and four-vector meson vertices into their
hidden local symmetry model, Alvarez-Ruso and Koch~\cite{Alvarez-Ruso:2002ib} found that the $\phi$ meson mean
free path in nuclear media is smaller than that usually
estimated. Ishikawa $et~al.$~\cite{Ishikawa:2004id} presented new
data on near-threshold $\phi$ photoproduction on several nuclear
targets. They found that the cross section between $\phi$ meson and
a nucleon, $\sigma_{\phi N}$, is equal to $35^{+17}_{-11}$ mb, which appears to be much larger than previous expectations although the experimental uncertainty is very large~\cite{Behrend:1978fr}. Meanwhile, Sibirtsev $et~al.$~\cite{Sibirtsev:2006yk} presented a new analysis of existing
$\phi$ photoproduction data and found $\sigma_{\phi N}$ $\sim10$
mb. Thus, $\sigma_{\phi N}$ in heavy-ion collisions is still
unclear.

The measurement of  collective radial flow (represented by $\langle p_T
\rangle$) probes the equation of state of matter produced in nuclear
collisions~\cite{Levai:1991be, Stocker:2005de}. Strong radial flow
has been observed for many particles such as $\pi$, $K$, and
$p(\bar{p})$~\cite{Adams:2003xp}. If the $\phi$ meson has indeed
small hadronic rescattering cross sections and decouples early from
the collision system, contributions to the radial flow of the $\phi$
meson will be mostly from the partonic stage instead of the hadronic
stage. Thus the $\phi$ meson may have a significantly smaller radial
flow than other hadrons with similar mass, such as the proton,
especially in central heavy-ion collisions. Therefore, $\phi$ mesons
may carry information about the conditions of nuclear collisions
before chemical freeze-out. Thus it is important to experimentally
measure and compare the freeze-out properties of the $\phi$ to other
hadrons as a function of centrality and collision species. A
comprehensive set of measurements will shed light on the
characteristics of $\phi$ meson production and the evolution of the
collision system.

The elliptic flow parameter $v_{2}$ is a good tool for studying the
system formed in the early stages of high energy collisions at RHIC.
It has been found that at low $p_{T}$ ($0<p_{T}<2$~GeV/$c$), the
dependence of $v_{2}$ on particle mass~\cite{Adler:2001nb,
Adler:2003kt, Adler:2002pb} is consistent with hydrodynamic
calculations in which local thermal equilibrium of partons is
assumed~\cite{Adams:2005dq,Ollitrault:1992bk, Sorge:1998mk, Huovinen:2001cy,
Teaney:2000cw}. This observation indicates that thermally
equilibrated partonic matter may have been created at RHIC. However,
at intermediate $p_{T}$ ($2<p_{T}< 5$~GeV/$c$), the measured $v_{2}$
for various hadrons seems to depend on the number of constituent
quarks in the hadron rather than its mass, consistent with the
results from coalescence/recombination models~\cite{Lin:2002rw,
Voloshin:2002wa, Greco:2003xt, Fries:2003vb, Fries:2003prc,
Molnar:2003ff}. Since $\phi$ is a meson but has a mass close to
that of the proton and $\Lambda$, the measurement of the $\phi$
meson elliptic flow will provide a unique tool
for testing the above statement. 

Current measurements of various hadrons by STAR ($\Lambda$, $p$,
$K^0_S$, $K(892)^{*}$, $h^{\pm}$, etc.) show that the nuclear
modification factor $R_{cp}$ for baryons differs from that of
mesons~\cite{Adams:2003am, Adams:2005kstar}, consistent with the
prediction of quark coalescence/recombination
models~\cite{Das:1977cp, Fries:2003vb,Fries:2003prc, Hwa:2003bn}.
Because of the value of the $\phi$ meson mass, a comparison of
$R_{cp}$ for the $\phi$ with these previous measurements will
conclusively determine if the observed difference is driven by
particle mass or particle type.

In previous studies, the production of the $\phi$ meson has been
measured in Au+Au collisions at $\sqrt{s_{_{NN}}}=$
130~GeV~\cite{Adler:2002xv} and in Au+Au and $p+p$ collisions at
$\sqrt{s_{_{NN}}}=$ 200~GeV~\cite{Adams:2004ux} at RHIC. The Au+Au
200~GeV data presented in this paper were taken from the year 2004
run, where the number of events is approximately ten times larger
than the previously reported number for the Au+Au run in year
2001~\cite{Adams:2004ux}. The data from the year 2004 run contain  the data reported in ref.~\cite{Abelev:2007phi, Blyth:2008phd}. It has been found that the results from the two runs are consistent with each other.  In this paper, we present
systematic measurements of $\phi$ meson production over a broad
range of collision energies and system sizes, including Au+Au
collisions at $\sqrt{s_{_{NN}}}=62.4$, 130~\cite{Adler:2002xv},
and 200~GeV, and $p+p$~\cite{Adams:2004ux} and $d$+Au
collisions at $\sqrt{s_{_{NN}}}=200$~GeV from the STAR experiment.
In Sec. II, we briefly introduce the STAR detector
and discuss our analysis method (event-mixing technique) in detail.
In Sec. III, we present the measurement of $\phi$ meson invariant
mass distributions (III~A), transverse mass $m_T$) spectra (III~B,
particle ratios (III~C), nuclear modification factors (III~D) and
the elliptic flow parameter $v_{2}$ (III~E); we also discuss the
physics implication of each of these results. A summary and
conclusions are presented in Sec. IV.


\begin{center}
\textbf{II.~DATA ANALYSIS}
\end{center}

\begin{center}
 \textbf{A.~Experimental Setup}\end{center}

The STAR detector~\cite{Ackermann:2002ad} consists of several
subsystems in a large solenoidal analyzing magnet. We discuss
here the subdetectors used in the analyses relevant to this paper. With its axis aligned
along the beam direction, the time projection chamber (TPC)~\cite{Anderson:2002ad}
 is the main tracking device for
charged particles, covering a pseudorapidity range $|\eta|\leq1.8$
and providing complete azimuthal coverage. The entire TPC is
located inside a solenoidal magnet, and data are taken at the maximum
magnetic field $|B_{z}|$ = 0.5 Tesla, where the $z$ axis is
parallel to the beam direction. Radial-drift TPCs (FTPCs)~\cite{Ackermann:2002ftpc} 
are also installed to extend
particle tracking into the forward and backward regions ($2.5<|\eta| <4.0$).
Surrounding the TPC is the central trigger barrel (CTB)~\cite{Bieser:2002triger}, which is a scintillator
counter array whose analog signal is sensitive to the total
charged particle multiplicity with coverage $|\eta|\leq1.0$. A
pair of beam-beam counters (BBCs) at $3.3<\eta<5.0$ and a pair of
zero degree calorimeters (ZDCs)~\cite{Adler:2002zdc}
at $\theta< 2$~mrad are located on either side of the collision
region along the beam line, and are used to provide event triggers
for data taking. A more detailed description of the STAR
detector can be found in Ref.~\cite{Ackermann:2002ad} and
references therein.

\begin{center}
\textbf{B.~Event selection}\end{center}

\begin{flushleft}
\textbf{\texttt{1.~$Trigger~selection$}}\end{flushleft}

The Au+Au data used in this analysis were taken with two different
trigger conditions. One was a minimum-bias (MB) trigger requiring
only a coincidence between both ZDCs. The other was a central
trigger additionally requiring both a large analog signal in the
CTB indicating a high charged particle multiplicity at
midrapidity and a small ZDC signal. The ZDCs measure
beam-velocity neutrons from the fragmentation of colliding nuclei
and were used as the experimental level-0 trigger for selecting
$d$+Au and Au+Au collisions for their respective runs. For $p+p$ data
taking, the BBCs were used as trigger detectors. The central
trigger corresponds to approximately the top $15\%$ and $12\%$ of
the measured cross section for Au+Au collisions at 130~GeV and
200~GeV, respectively. Data from both the MB and central triggers
were used for this analysis. For the Au+Au 62.4~GeV data set, only
MB triggered events were used. For $p+p$ collisions at 200~GeV, the
MB trigger was used in the analysis. It was based on a
coincidence between the two BBCs. The BBCs are sensitive
only to the non-single diffractive (NSD) part (30~mb) of the $p+p$
total inelastic cross section (42~mb)~\cite{Adams:2004ux}.

\begin{flushleft}
\textbf{\texttt{2.~$Vertex~cuts~and~centrality~selection$}}
\end{flushleft}

The longitudinal $z$ position of the interaction point is determined on-line by the measured time difference of the two ZDCs's signals. A cut on the $z$ position of the interaction point is applied on-line for
all data sets (except $p+p$) in order to maximize the amount of useful data for physics analysis, since events
with primary vertices far away from the center of the TPC have a significant non-uniform acceptance. In the off-line
data analysis further cuts are applied on the $z$ position of the reconstructed primary vertex ($V_Z$), to
ensure nearly uniform detector acceptance. These cuts are listed in Table~\ref{tab:vertexz}.
\begin{table}
\begin{tabular*}{\hsize}{l|c|c|c|c|c|l}
 \hline
 \hline
{System} & {$\sqrt{s_{_{NN}}}$} & {Trigger} & {$|V_Z|$} & {Centrality} & {Events}  & {Year}\\
{} & {(GeV)} & {} & {(cm)} & {} & {} & {}\\
\hline
  Au+Au & 62.4  & MB & $\leq$ 30 & 0-80\%   & $6.2\times 10^6$ & 2004\\
  Au+Au & 130 & MB & $\leq$ 80 & 0-85\%   & $7.6\times 10^5$ & 2000\\
  Au+Au & 130 & Central  & $\leq$ 80 & 0-11\%   & $8.8\times 10^5$ & 2000\\
  Au+Au & 200 & MB & $\leq$ 30 & 0-80\%   & $1.4\times 10^7$ & 2004\\
  Au+Au & 200 & Central  & $\leq$ 30 & 0-5\%   & $4.8\times 10^6$ & 2004\\
  $p+p$   & 200 & MB & $\leq$ 50 & MB (NSD) & $6.5\times 10^6$ & 2002\\
  $d$+Au  & 200 & MB & $\leq$ 50 & MB   & $1.4\times 10^7$ & 2003\\
\hline \hline

\end{tabular*}
\caption{Data sets used in the analysis. Cuts on $V_Z$, the selected centrality ranges and the final number of events included in the analysis after all cuts/selections are also shown.} \label{tab:vertexz}
\end{table}

To define the collision centrality for the Au+Au data, the raw
charged hadron multiplicity distribution in the TPC within a
pseudo-rapidity window $|\eta| \le$ 0.5 ($|\eta| \le$ 0.75 was
used for the Au+Au 130~GeV data) was divided into several bins.
Each bin corresponds to a certain fraction of the total inelastic
cross section~\cite{Abelev:2007}. For the $d$+Au data, the raw
charged hadron multiplicity in the east (Au-direction) FTPC (-3.8 $< \eta <$ -2.8) was used for the centrality definition to avoid auto-correlation between centrality and the measurements of charged particles at midrapidity in the TPC~\cite{Abelev:2007}. We defined four centrality bins for the Au+Au 62.4~GeV data (0-20\%, 20-40\%, 40-60\%, 60-80\%), three
centrality bins for the Au+Au 130~GeV data (0-11\%, 11-26\%,
26-85\%), nine centrality bins for the Au+Au 200~GeV data (0-5\%,
0-10\%, 10-20\%, 20-30\%, 30-40\%, 40-50\%, 50-60\%, 60-70\%,
70-80\%) and three centrality bins for the $d$+Au 200~GeV data
(0-20\%, 20-40\%, 40-100\%). The 80-100\% most peripheral Au+Au
collision data were not used because of the rapidly decreasing
trigger and vertex finding efficiencies for low multiplicity
events. Table~\ref{tab:vertexz} lists the data sets used along
with centrality selections and final numbers of events after these
cuts. Note that the elliptic flow ($v_{2}$) measurement only from
200~GeV MB Au+Au data is presented in this paper, as the
statistics for the $v_{2}$ analysis is  not sufficient for the
62.4 and 130~GeV Au+Au data.

\begin{center}
\textbf{C.~Track selection and particle identification}\end{center}
\begin{flushleft}
\textbf{\texttt{1.~$Track~selection$}}\end{flushleft}

Several quality cuts were applied to ensure selection of good tracks. During the TPC track reconstruction,
a charged track was extrapolated back to the beam line by using the reconstructed helix parameters. If the
distance of closest approach (DCA) of the track to the event vertex was less than 3~cm and the track had at least
ten hit points in the TPC, the reconstructed track was labeled as a primary track. The helix parameters for primary
tracks were then refitted by requiring that the helix pass through the primary vertex location. This procedure
improved the momentum resolution of tracks. Since the $\phi$ meson has a very short
lifetime, it decays at the primary vertex position. Thus only primary tracks were used
for the $\phi$ meson analysis. As a systematic check, the DCA selection for primary tracks was changed from 3 cm
to 1 cm. The differences in the results were small and were included in the  estimate of systematic uncertainties. Tracks with transverse momentum less than 0.1 GeV/$c$ were not used, as their combined acceptance and efficiency becomes very small.  Each track included in the $\phi$
analysis was required to have at least 15 hit points out of 45 used in the fitting of the tracks helix parameters. The ratio of the
number of space points used in the track reconstruction to the maximum possible number of hit points was
required to be greater than $55\%$ to avoid split tracks where a real track is reconstructed in two or more segments. A
pseudo-rapidity cut $|\eta|<1.0$ ($|\eta|<1.1$ for Au+Au 130~GeV data) was applied to
select tracks that are well within the TPC acceptance.

\begin{flushleft}
\textbf{\texttt{2.~$Kaon~selection$}}\end{flushleft}

\begin{figure}[t]
\centering
\includegraphics[width=.50\textwidth]{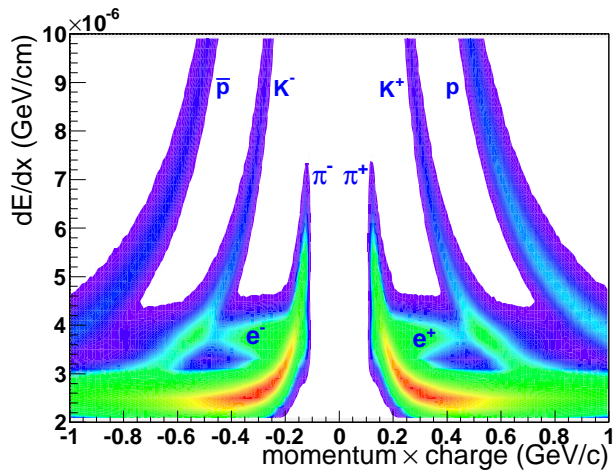}
\caption{\footnotesize(Color online) Measured $\langle dE /dx
\rangle$ vs momentum $\times$ charge of reconstructed tracks in
the TPC. The figure is generated from Au+Au 62.4~GeV data.}
\label{dedx}
\end{figure}

Particle identification (PID) was achieved by correlating the ionization energy loss ($dE/dx$) of charged
particles in the TPC gas with their measured momentum. The measurement of mean $dE/dx$ was achieved by averaging
the measured $dE/dx$ samples along the track after truncating the top $30\%$.
The measured $\langle dE/dx\rangle$ versus momentum curve is reasonably well described by the Bethe-Bloch
function~\cite{Yao:2006ee} smeared with the detector's resolution (note that the Bichsel function was used to fit the
$\langle dE/dx\rangle$ plot in Au+Au 200 GeV from the year 2004 run~~\cite{Bichsel:2006de}). The $n_{\sigma}$ values for kaons are calculated via
\begin{equation}
n_{\sigma} = \frac{1}{R}\log{\frac{\langle
dE/dx\rangle_{measured}}{dE/dx_{expected}}},
\label{nsigma}
\end{equation}
where $\langle dE/dx\rangle_{measured}$ and $dE/dx_{expected}$ are
$\langle dE/dx \rangle$ measured by TPC and calculated analytically,
respectively, and $R$ denotes the $dE/dx$ resolution of the track which is found to range between $6\%$ and $10\%$. $R$ is determined experimentally and depends on the event
multiplicity and the number of $dE/dx$ samples from the track used to calculate the mean value. Tracks within
2$\sigma$ of the kaon Bethe-Bloch curve were selected as kaon
candidates. Figure~\ref{dedx} presents the measured $\langle
dE/dx\rangle$ versus momentum $\times$ charge in Au+Au
collisions at 62.4~GeV. Table~\ref{tab:trackselection} lists all the track cuts applied in the analysis.

\begin{table}[htbp]
\centering
\begin{tabular*}{240pt}{lc}
\hline \hline
{Cut parameter} & {Value}\\
\hline {Track DCA (cm)} & {$< 3$}\\
{Track $N_{Fit}$} & {$\geq15$}\\
{Track $N_{Fit}$/$N_{Max}$} & {$>0.55$}\\
{Track momentum (GeV/$c$)} & {0.1 $<p<$ 10}\\
{Track transverse momentum (GeV/$c$)} & {0.1 $<p_T<$ 10}\\
{Kaon $dE/dx$} & {$|n_{\sigma}|<2.0$ (for kaon)}\\
{$\phi$ candidate's $\delta\!\!-\!\!dip\!\!-\!\!angle$ (radians)} & {$> 0.04$}\\
{$\phi$ candidate's rapidity} & {$|y|<0.5$ (for spectra) }\\
{}& {$|y|<1.0$ (for $v_{2}$) }\\
\hline \hline
\end{tabular*}

\caption{Track cuts used in the analysis, where $N_{Fit}$ and $N_{Max}$ represent the number of fitted hits and the maximum number of hits for TPC tracks, respectively.}

\label{tab:trackselection}
\end{table}

Note that from the $\langle dE/dx\rangle$ measurement, kaons
cannot be clearly separated from pions above $p\sim$ 0.6~GeV/$c$ and from
protons/antiprotons above $p\sim$ 1.1~GeV/$c$. Also note that the electron and positron $dE/dx$ bands cross the bands
for pions, kaons, and protons/antiprotons. Therefore selected kaon
candidates are contaminated by electrons/positrons, pions and
protons/antiprotons varying with $p$. Contamination by
these charged particles in the kaon sample brings in additional
real correlations (such as particle decays) which cannot be
subtracted by the event-mixing method. We have varied the $n_{\sigma}$ cut for kaon to investigate the
efficiency and combinatorial background dependence of the $\phi$
signal extraction on this cut. The resulting systematic
uncertainties have been included in the estimate of the total
systematic errors.

\begin{center}
\textbf{D.~Event mixing and raw yield extraction}\end{center}

The $\phi$ meson signal was generated by pairing all $K^+K^-$ tracks
from the same event that passed the selection criteria and by then
calculating the invariant mass $m_{inv}$ for all possible $K^+K^-$ pairs. As random combinations of $K^+K^-$ pairs are
dominant in this process, the resulting same-event invariant mass
distribution contains the $\phi$ meson signal on top of a large
combinatorial background. An event-mixing
technique~\cite{L'Hote:1992qz,Drijard:1984pe} was applied to
calculate the shape of the combinatorial background, where the
invariant mass was calculated by pairing two kaons from two
different events with same primary vertex and multiplicity bins (mixed event). Ideally, since it combines two
different events, the mixed-event distribution contains
everything except the real same-event correlations.

The STAR TPC has symmetric coverage about the center of the
collision region. However, variations in the acceptance occur, 
since the collision vertex position may change considerably
event-by-event. This variation in the collision vertex position
gives rise to a nonstatistical variation in the single-particle
phase-space acceptance, which would lead to a mismatch between the
mixed-event and same-event invariant mass distributions. This
mismatch would prevent the proper extraction of the $\phi$ meson
signal. By sorting events according to their primary vertex $V_Z$
position and performing event-mixing only among events in the same
vertex bin, the mismatching effect is minimized. In this analysis,
events were divided into $V_Z$ bins that  were 6~cm wide ($V_Z$
resolution is $\sim$~0.3~mm) for event-mixing. To further
improve the description of the background, two events were only
mixed if they had similar event multiplicities. These requirements
ensure that the two events used in mixing have similar event
structures, so the mixed-event invariant mass distribution can
better represent the combinatorial background in the same-event
invariant mass distribution. Consistent results were obtained when we
constructed the background distribution using like-sign pairs
from the same event.

To reduce statistical uncertainty in the mixed event,
each event was mixed with five to ten other events (depending on
the collision system). To extract the $\phi$ meson signal, the
mixed-event and same-event $K^+K^-$ invariant mass distributions
were accumulated and the mixed-event distribution normalized to
the same-event distribution in the region above the $\phi$ mass,
$1.04 < m_{inv} < 1.06$ GeV/$c^2$, and subtracted in each
$p_T$ and $y$ (rapidity) bin for every collision centrality. We
varied the normalization region and normalization factor to
estimate the systematic uncertainty on the normalization, and the
estimated uncertainty was included in the quoted total systematic
uncertainty.

Despite the requirements for mixing events described above, a
residual background remains over a broad mass region in the
subtracted invariant mass distribution. This is due to an
imperfect description of the combinatorial background and the fact
that the mixed event cannot account for the real correlated
background from decay pairs due to Coulomb interactions, photon
conversions ($\gamma \rightarrow e^+e^-$), and  particle decays
such as $K^{0*}\rightarrow K^+\pi^-$, $\rho^0\rightarrow \pi^+\pi^-$,
$K_S^0\rightarrow \pi^+\pi^-$, and $\Lambda\rightarrow
p\pi^-$~\cite{Yamamoto:2001da}. For example, when both pions from
a $K_S^0$ decay are misidentified as kaons, the real correlation
from decay will remain in the same-event as a broad distribution,
but will not be reproduced by the event-mixing method.

\begin{figure}[t]
\centering
\includegraphics[width=.50\textwidth]{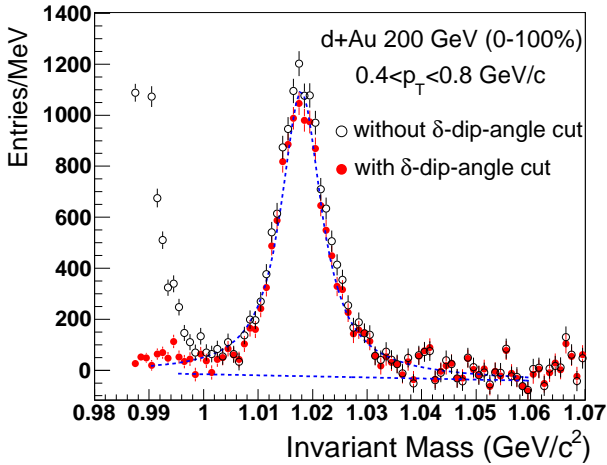}
\caption{\footnotesize(Color online) Background-subtracted
invariant mass distributions at $0.4<p_{T}<0.8$ GeV/$c$ in $d$+Au
200~GeV collisions (0-100\%) with (solid points) and without (open
points) the $\delta$-dip-angle cut. The dashed curves show a
Breit-Wigner (see the text for details) + linear background
function fit to the case with the $\delta$-dip-angle cut.}
\label{dipangle}
\end{figure}

Due to contamination of electrons/positrons in the selected kaon
sample, the $K^+K^-$ invariant mass distribution contains residual
background near the threshold from correlated $e^+e^-$ pairs,
mainly from photon conversions ($\gamma \rightarrow e^++e^-$). The
$\delta$-dip-angle between the photon converted electron and
positron is usually quite small. The $\delta$-dip-angle is calculated from
\begin{equation}
\delta\!\!-\!\!dip\!\!-\!\!angle=cos^{-1}[\frac{p_{T1}p_{T2}+p_{z1}p_{z2}}{p_1p_2}],
\label{eq:DeltaDipAngle}
\end{equation}
where $p_1$, $p_2$, $p_{T1}$, $p_{T2}$, $p_{z1}$, $p_{z2}$ are
momentum and transverse and longitudinal momentum components of the
two tracks; this represents the opening angle of a pair in the $p_z$-$p_T$ plane. We required the $\delta$-dip-angle to be greater than
0.04 radians. This cut was found to be very effective in removing
the photon conversion background while only reducing the $\phi$
reconstruction efficiency by $\sim$ 12\%. Figure~\ref{dipangle}
shows two  background-subtracted invariant mass distributions with
and without the $\delta$-dip-angle cut. One can see that the peak
from photon conversion($m_{inv}\leq 1.0$ GeV/$c^{2}$) is very
effectively removed by this cut.

\begin{figure*}[htb]
\hspace*{-21mm}
\begin{minipage}[t]{69mm}
\includegraphics[width=68mm, height=90mm]{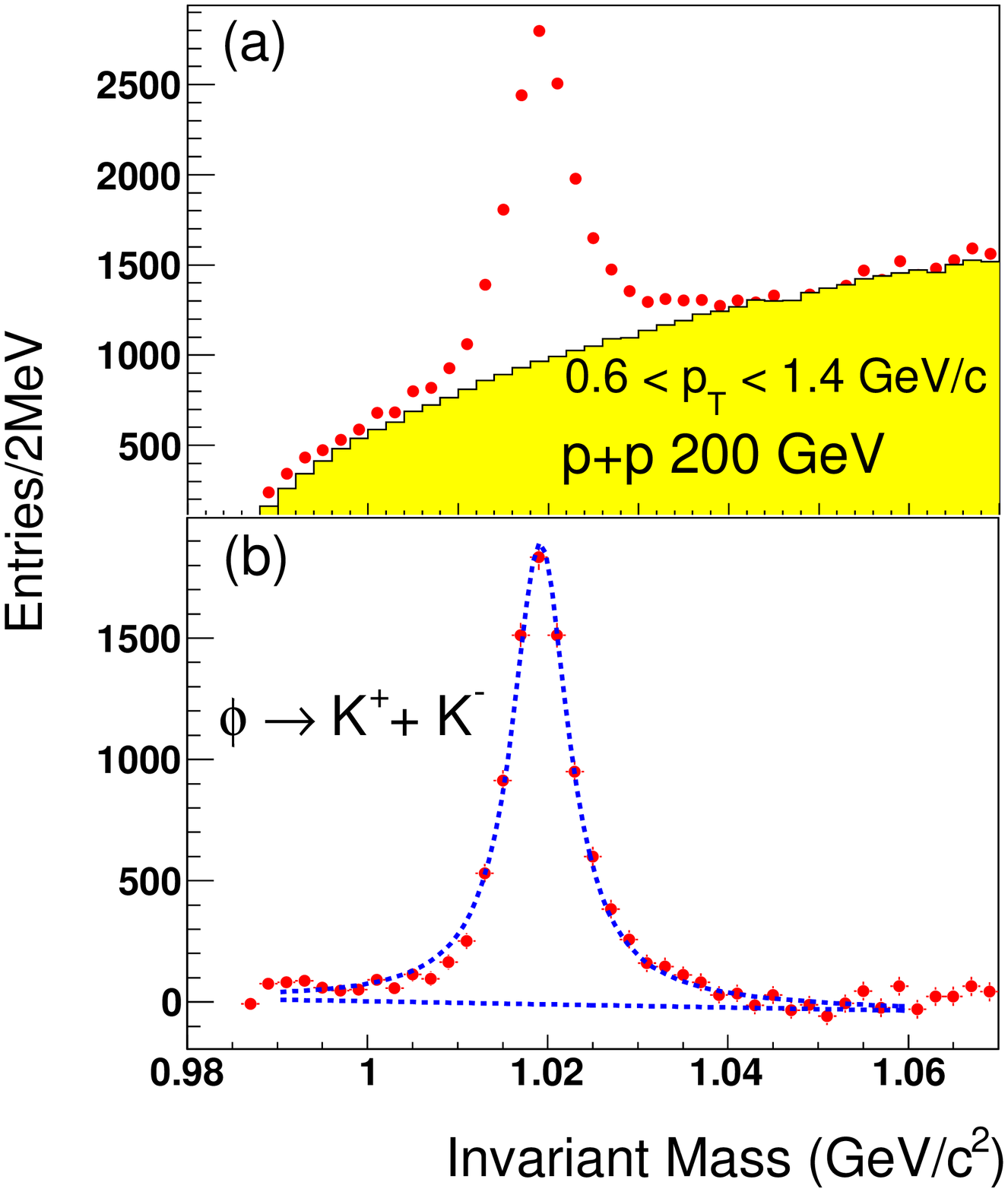}
\end{minipage}
\hspace{-7mm}
\begin{minipage}[t]{69mm}
\includegraphics[width=68mm, height=90mm]{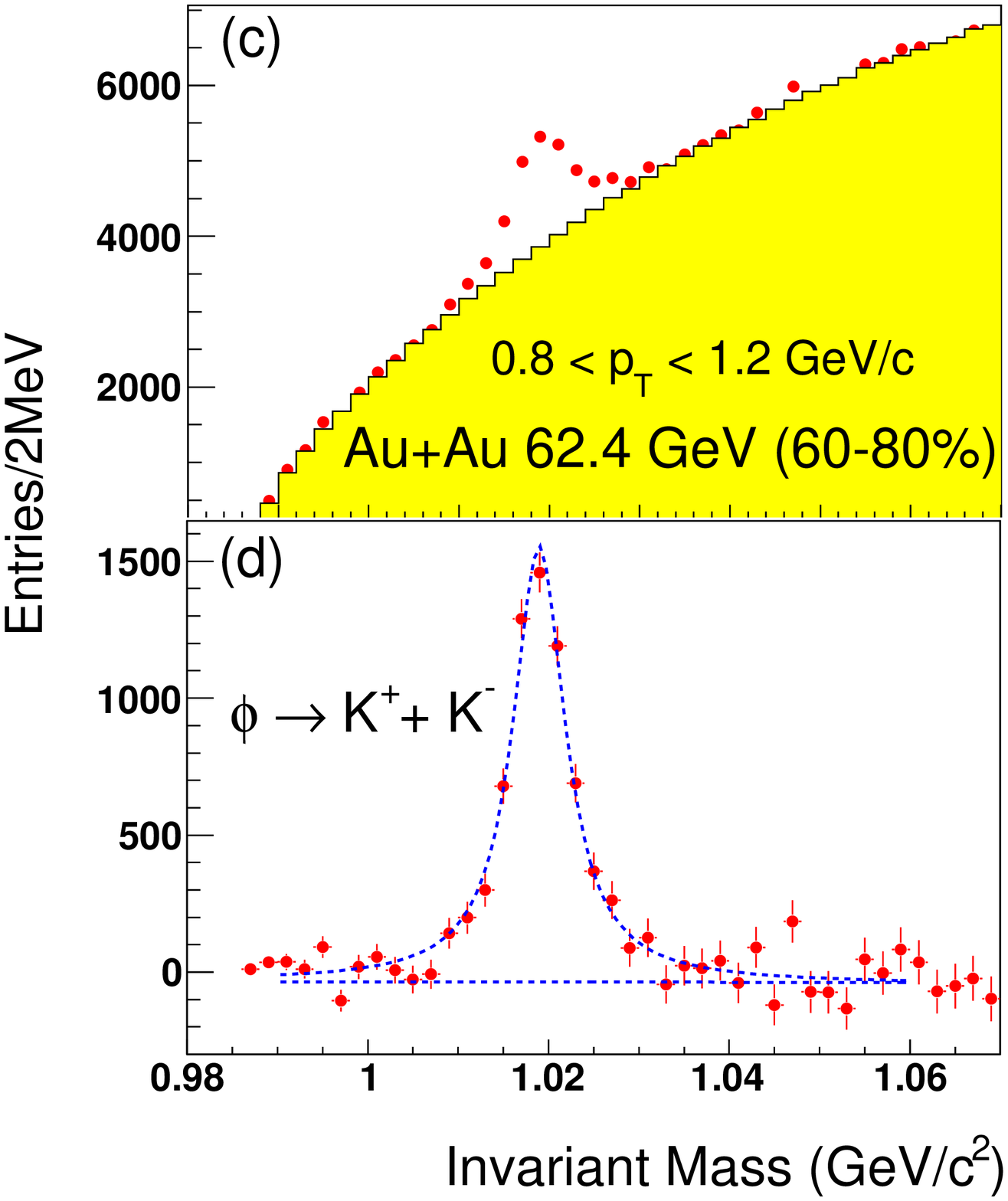}
\end{minipage}
\hspace{-7mm}
\begin{minipage}[t]{69mm}
\includegraphics[width=68mm, height=90mm]{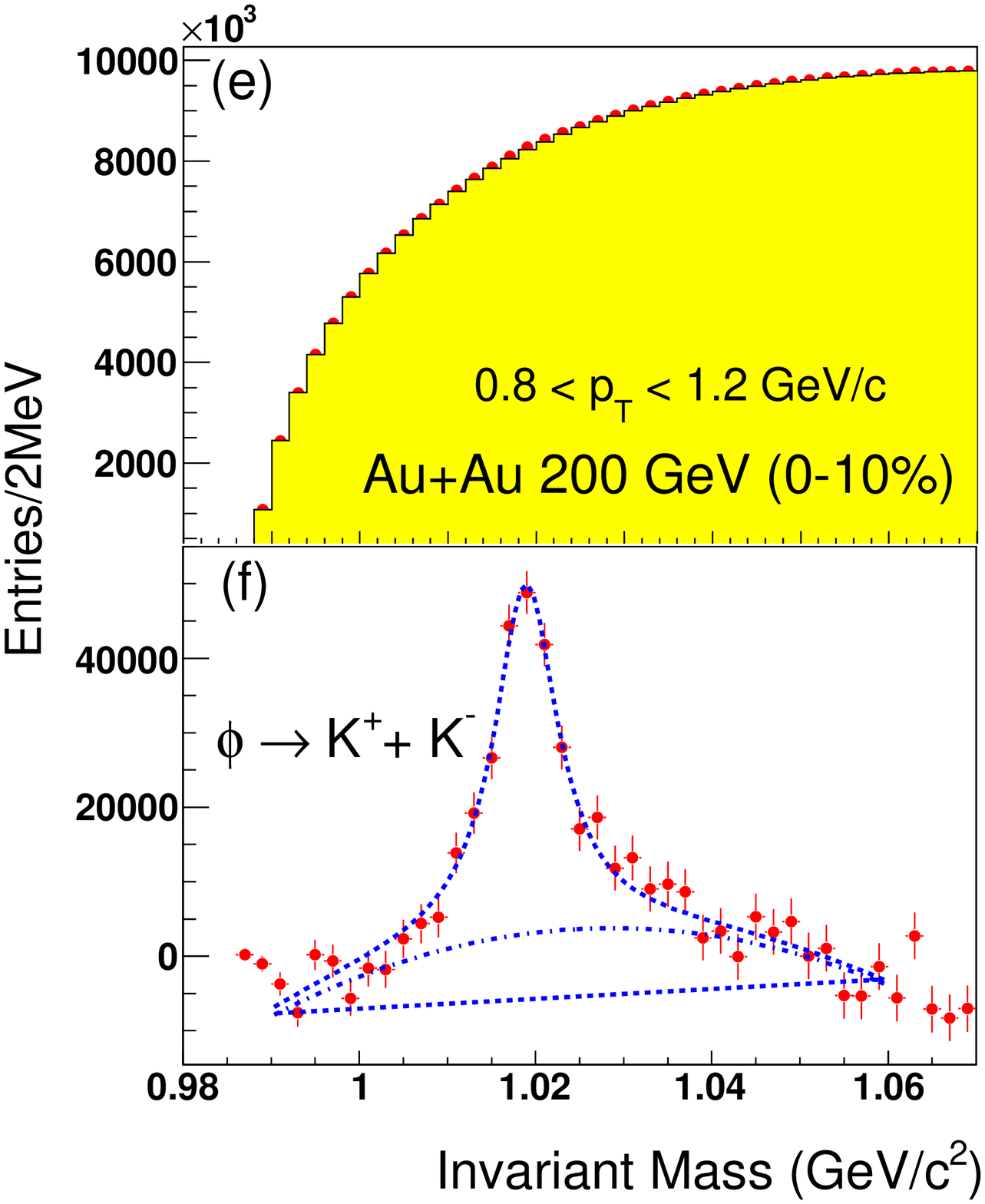}
\end{minipage}
\hspace*{-17mm}
\caption{\footnotesize(Color online) Upper panels: same-event (full
points) and mixed-event (solid line) $K^+K^-$ invariant mass distributions
at $0.6<p_{T}<1.4$ GeV/$c$ in $p+p$ 200~GeV collisions (a), $0.8<p_{T}<1.2$ GeV/$c$ in Au+Au 62.4~GeV collisions (60-80\%) (c) and $0.8<p_{T}<1.2$ GeV/$c$ in Au+Au 200~GeV collisions (0-10\%) (e). Lower panels: the corresponding $\phi$ meson mass peaks after subtracting the
background. Dashed curves show a Breit-Wigner + linear background function fit in (b),  (d). In (f), both linear and quadratic backgrounds are shown as dashed and dot-dashed lines, respectively. } \label{signal_get}
\end{figure*}

Figure~\ref{signal_get} shows the $K^+K^-$ invariant mass
distributions for $p+p$ collisions at 200~GeV [(a) and
(b)], 60-80\% Au+Au collisions at 62.4~GeV [(c) and
(d)] and 0-10\% Au+Au collisions at 200~GeV [(e) and (f)]. Solid
circles in the upper panels are same-event pairs, whereas the
histograms are from mixed-event pairs. The $\phi$ meson peak is clearly
visible for $p+p$ 200~GeV and Au+Au 62.4~GeV (60-80\%) in Figs.~\ref{signal_get}(a) and (c) 
before background subtraction, but not for Au+Au
200~GeV (0-10\%) [Fig.~\ref{signal_get}(e)] due to its smaller signal significance. However, after background subtraction, the
$\phi$ mass peak can be seen clearly for all data sets.
The lower panels in Fig.~\ref{signal_get} show the mixed-event
background-subtracted $\phi$ invariant mass distributions. Raw
yields for the $\phi$ meson were determined by fitting the
background-subtracted $m_{inv}$ distribution with a Breit-Wigner
function superimposed on a linear (or polynomial) background
function
\begin{equation}
\frac{dN}{dm_{inv}}=\frac{A\Gamma}{(m_{inv}-m_{0})^{2}+\Gamma^{2}/4}+B(m_{inv}), \label{equ:BW}
\end{equation}
where $A$ is the area under the peak corresponding to the number of $\phi$ mesons, $\Gamma$ is the full width at half maximum (FWHM) of the peak,
and $m_{0}$ is the resonance mass position. $B(m_{inv})$ denotes a linear [$B(m_{inv})=p_{0}+p_{1}m_{inv}$, shown by a dashed line in Figs.~\ref{signal_get}(b), (d) and (f)] or polynomial [$B(m_{inv})=p_{0}+p_{1}m_{inv}+p_{2}m_{inv}^{2}$, shown by a  dot-dashed line in Fig.~\ref{signal_get}(f)] residual background function. The parameters $p_{0}$, $p_{1}$ and $p_{2}$ of $B(m_{inv})$ and $A$, $m_0$ and $\Gamma$ are free parameters.

\begin{center}
\textbf{E.~Efficiency correction}\end{center}

\begin{figure}[htb]
\centering
\includegraphics[width=.50\textwidth]{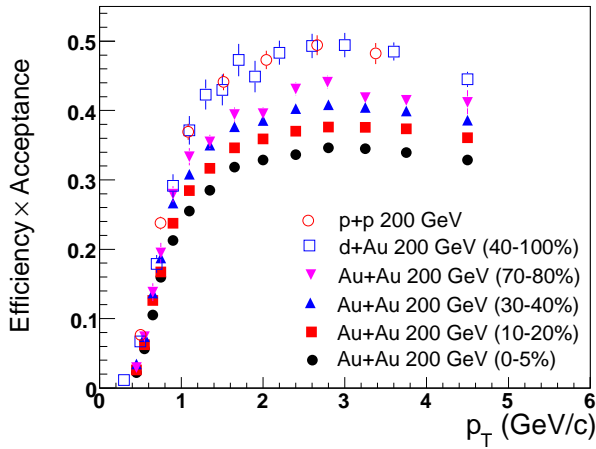}
\caption{\footnotesize(Color online) Reconstruction efficiency including acceptance of $\phi$ meson as a function of $p_{T}$ in several centrality bins of Au+Au,  $d$+Au and $p+p$ 200~GeV collisions.} \label{efficiecy}
\end{figure}

The $\phi$ acceptance and reconstruction efficiency were calculated
using an embedding technique, in which simulated tracks were embedded
into real events. The number of embedded simulated tracks is
approximately 5\% of the measured multiplicity of the real event.
The $\phi$ meson decay ($\phi\rightarrow K^{+}K^{-}$) and the
detector responses were simulated by the GEANT program
package~\cite{Carminati:1993GEANT} and the simulated output signals
were embedded into real events before being processed by the
standard STAR event reconstruction code. Embedded data were then
analyzed to calculate tracking efficiencies and detector acceptance
by dividing the number of reconstructed $\phi$ by
the number of input $\phi$ in the desired kinematic regions.
Figure~\ref{efficiecy} shows examples of correction factors
(tracking efficiency $\times$ acceptance) for our analysis as a
function of $\phi$ meson $p_{T}$ for selected centrality bins for
Au+Au 200~GeV, $d$+Au 200~GeV and $p+p$ 200~GeV collisions. It can be
seen that the overall correction factors increase from a few percent
at low $p_T$ to over 30\% at high $p_T$. Low efficiency at low $p_T$
is mainly due to poor acceptance of the daughter tracks. The
efficiency is lower in more central collisions because of the increasing
occupancy in the TPC~\cite{Abelev:2007}.

\begin{center}
\textbf{F.~Vertex finding and trigger efficiency
correction}\end{center}

For $p+p$ and $d$+Au data, the trigger efficiency is less than
100\%~\cite{Adams:2006nd}. The MB trigger for $p+p$ data was found to
trigger $\sim$ 87\% of $p+p$ NSD events. For $d$+Au data, the trigger
efficiency was found to be $\sim$ 95\%.
These trigger efficiencies were used to normalize the measured yield
and the corresponding uncertainties are added to the total
systematic errors for $p+p$ and $d$+Au data. The MB trigger efficiency
for Au+Au data is essentially 100\% ~\cite{Abelev:2007}.

It was found that the event vertex finding efficiency, which is the
fraction of events having reconstructed vertices, drops rapidly for
low multiplicity events~\cite{Abelev:2007}. For $d$+Au collisions, the vertex efficiency was 88\%
for the most peripheral bin (40-100\%), 93\% for the MB events and 100\% for the central and middle central bins (0-20\% and
20-40\%). For Au+Au collisions, the vertex finding efficiency was 99.9\% and $p_T$ independent~\cite{Jingguo:2006da} due to the
increased track multiplicity in those collisions. However the overall vertex finding efficiency was found to be 98.8\% for
the MB $p+p$ data by applying an additional BBC time difference selection (i.e., the NSD requirement), and the effect of correction was negligible in this analysis.

\begin{center}
\textbf{G.~$v_{2}$ measurement}\end{center}

\begin{flushleft}
\textbf{\texttt{1.~$Reaction~plane~method$}}\end{flushleft}

We employed the STAR standard reaction plane method as
described in Refs.~\cite{Ackermann:2000tr, Poskanzer:1998yz}, which uses
a Fourier expansion to describe particle emission with respect to
the reaction plane angle, that is
\begin{equation}
E\frac{d^3N}{d^3p}=\frac{1}{2\pi}\frac{d^2N}{p_tdp_tdy}\left(1+
\sum_{n=1}^{\infty}2v_n\cos[n(\varphi-\Psi_r)]\right),
\label{eq:v2}
\end{equation}
where $\Psi_r$ is the real reaction plane angle and $\varphi$ is
the particle's azimuthal angle. The coefficient $v_{2}$ in the
second-order term of the expansion is the dominant part and is
called the second harmonic anisotropic flow parameter, or elliptic flow.

The real reaction plane angle $\Psi_r$ is not known, but can be estimated experimentally~\cite{PRC-STAR-Charged-v2}. In our analysis, the estimated reaction plane angle from the second-order harmonic $(\Psi_2)$ was used. This has a finite resolution due to a limited number of particles available in each event and a different event-by-event $v_{2}$,  which is used for the estimation. The estimated reaction plane resolution was used to correct the observed $v_2^{obs}$ to obtain the final estimation of $v_{2}$.

The event plane angle $\Psi_2$ was calculated by the equation
\begin{equation}\label{EP1}
\Psi_2=\frac{1}{2}\tan^{-1}\bigg(\frac{\sum_{i}w_{i}\sin(2\varphi_{i})}{\sum_{i}
w_{i}\cos(2\varphi_{i})}\bigg),
\end{equation}
where the sums are over all charged particles used for reaction plane determination, $w_{i}$ and
$\varphi_i$ are the weight and azimuthal angle for the $i$th particle in a given event, respectively. The weights include both a $p_T$ weight and $\varphi$ weight. The $p_T$ weight was taken to be the particle $p_T$ up to 2.0~GeV/$c$ and
constant (2.0) above that~\cite{Poskanzer:1998yz}. The $\varphi$ weight was taken to be the reciprocal of the $\varphi$ distribution (normalized by the average entries) for all selected tracks. The autocorrelations were eliminated by excluding all kaon tracks used in the $\phi$ invariant mass
calculation from the reaction plane angle estimation~\cite{Jingguo:2006da}.

\begin{figure}[htb]
\centering
\includegraphics[width=.50\textwidth]{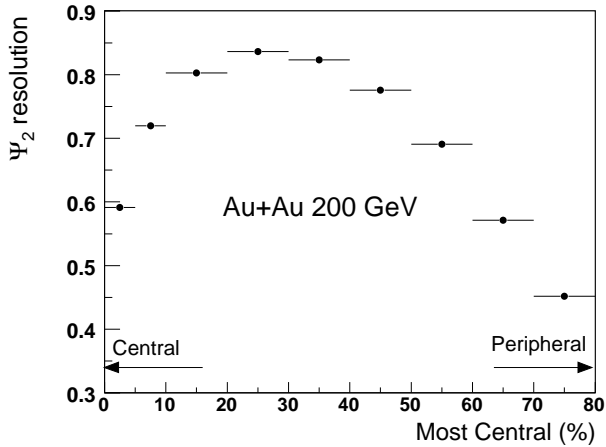}
\caption{\footnotesize Event plane $\Psi_2$ resolution as a
function of centrality in Au+Au 200~GeV collisions, where the vertical axis starts from 0.3 for clarity.}
\label{resolution}
\end{figure}

The reaction plane resolution was then calculated by
\begin{equation}
\langle\cos[2(\Psi_2-\Psi_r)]\rangle=C\langle
\cos[2(\Psi_2^a-\Psi_r)]\rangle, \label{eq:EPResolution}
\end{equation}
where $\Psi_2^a$ is the calculated reaction plane angle of the subevent, and $C$ is a constant calculated from the known multiplicity dependence of the resolution~\cite{Poskanzer:1998yz}. This resolution was determined by dividing each event into two random subevents, a and b, with equal multiplicities. The reaction plane resolution therefore corresponds to how accurately the event plane angle represents the real reaction plane; due to its the definition in Eq.~(\ref{eq:EPResolution}),  a value of unity indicates ideal resolution.  Figure~\ref{resolution} presents the reaction plane resolutions in different centrality bins for Au+Au 200~GeV collisions.

\begin{figure}[htb]
\centering
\hbox{\includegraphics[width=.50\textwidth]{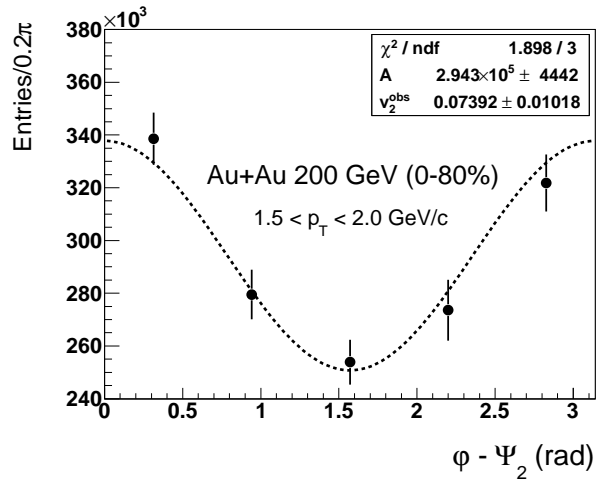}}
\caption{\footnotesize $\varphi-\Psi_2$ distribution for $\phi$ meson
at $1.5<p_{T}<2.0$~GeV/$c$ in Au+Au collisions (0-80\%) at 200~GeV.
The line is the fitting result. Error bars are statistical only.}
\label{v2cosfit}
\end{figure}

The combinatorial background in the $\phi$ invariant mass distribution
was also calculated by an event-mixing technique as described above.
To guarantee the mixed-event sample would represent the combinatorial background, an additional cut, the reaction plane angle difference between two mixed events,  was required to be less than 0.1$\pi$ rad in the event-mixing procedure.  After background subtraction,
$\phi$ meson yield was extracted in each $(p_T, \varphi-\Psi_2)$
bin. The yield distribution as a function of
$\varphi-\Psi_2$ was fitted by the function
\begin{equation}
A\bigg(1 + 2v_2^{obs}\cos[2(\varphi-\Psi_2)]\bigg), \label{eq:v2FitFunc}
\end{equation}
to extract the $v_2^{obs}$ value, where $A$ is a constant. A
typical result for the Au+Au data at 200~GeV is presented in
Fig.~\ref{v2cosfit}.

The measured $v_2^{obs}$ was then divided by the reaction plane
resolution to obtain the final $v_{2}$, i.e.,
\begin{equation}
v_2=\frac{v_2^{obs}}{\langle\cos[2(\Psi_2-\Psi_r)]\rangle}.\label{eq:realv2}
\end{equation}
Simulation studies have
found that the measured $v_{2}$ is about 7\% (relative to the real
$v_{2}$) lower than the real $v_{2}$ due to binning effects (five
bins in $\varphi-\Psi_2$); a correction has been made to the
measured $v_{2}$ to account for this effect.

\begin{flushleft}
\textbf{\texttt{2.~$Invariant~mass~method$}}\end{flushleft}

A new method, namely, the \emph{invariant mass method}, was also used to extract the
elliptic flow $v_{2}$ of the $\phi$ meson. The method was proposed in Ref.~\cite{Borghini:2004ra}, which decomposes  the anisotropic flow $v_{n}$ of a short-lived particle from  that of all possible daughter pairs as a function of invariant mass. For extracting the $v_{2}$ of the $\phi$ meson, it utilizes the fact that the $v_{2}$ of $K^+K^-$ pairs is composed of
the $v_{2}$ of the combinatorial background and the $v_{2}$ of the $\phi$ meson. Following the mixed-event background-subtraction procedure described in Sec. II~D,
the number of $K^+K^-$ pairs in each invariant-mass bin were counted, irrespective of the pair azimuth. Then
\begin{equation}
N_{K^+K^-}(m_{inv})=N_{\phi}(m_{inv})+N_{B}(m_{inv}), \label{eq:v2Fitinvmas1}
\end{equation}
where $N_{\phi}$ and $N_{B}$ are from the $\phi$ signal and the background, respectively.
Once  $N_{\phi}$ has been extracted via event-mixing and fitting the $\phi$ mass peak
with Eq.~(\ref{equ:BW}) for each $p_{T}$ bin as discussed in Section II~D,
$N_{B}$ can be obtained from Eq.~(\ref{eq:v2Fitinvmas1}).

The same-event $v_{2}$ for $K^+K^-$ pairs vs. invariant mass can be described by the function
\begin{equation}
v_{2}(m_{inv})=a(m_{inv})v_{2S} + [1-a(m_{inv})]v_{2B}(m_{inv}),
\label{eq:v2Fitinvmas}
\end{equation}
where $v_2(m_{inv})$ is the $v_{2}$ of same-event $K^+K^-$ pairs, $v_{2S}$ $\equiv$ $v_{2\phi}$ is the $v_{2}$ of
the $\phi$ meson, $v_{2B}$ is the effective $v_{2}$ of the combinatorial background and $a(m_{inv})$ =
$N_{\phi}(m_{inv})/N_{K^+K^-}(m_{inv})$ is the ratio of the $\phi$ signal to the sum of the background and $\phi$
signal. The reaction plane angle $\Psi_{2}$ was estimated in the same way as for the reaction
plane method described in the previous section. Therefore the two methods are not completely independent. $v_2(m_{inv})$ can then be calculated from the following
equation~\cite{Poskanzer:1998yz} for each $m_{inv}$ bin
\begin{equation}
v_{2}(m_{inv})=\langle \cos[2(\varphi_{KK}-\Psi_{2})] \rangle, \label{eq:v2Fitinvmas2}
\end{equation}
where $\varphi_{KK}$ is the azimuthal angle of the $K^+K^-$ pair.

\begin{figure}[t]
\centering
\includegraphics[width=.50\textwidth]{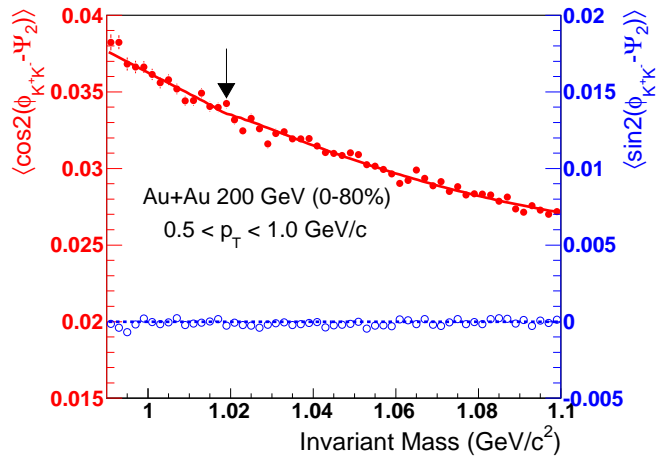}
\caption{\footnotesize(Color online) $\langle$cos[2$(\varphi_{K^+K^-}-\Psi_{2})] \rangle$ (full red points) and $\langle$sin[2$(\varphi_{K^+K^-}-\Psi_{2})] \rangle$ (open blue points) as a function of  $m_{inv}$ of $K^+K^-$ pairs at 0.5 $<$ $p_{T}$ $<$ 1.0 GeV/$c$ in Au+Au 200~GeV
collisions (0-80\%), where the solid curve is the result of fitting by
Equation~(\ref{eq:v2Fitinvmas}). The arrow shows the position of
$\phi$ invariant mass peak. The dash line shows zero horizontal line.} \label{v2invFit}
\end{figure}

Under the assumption that the background contribution to
$v_{2}(m_{inv})$ [the second part on right side of
equation~(\ref{eq:v2Fitinvmas})] is smooth as a function of
$m_{inv}$~\cite{Borghini:2004ra}, a polynomial function,
$p_{0}$+$p_{1}$$m_{inv}$+$p_{2}$$m^{2}_{inv}$, can be used to
parametrize the background $v_{2B}$ vs $m_{inv}$. $v_{2S}$ is
then obtained by fitting $v_{2}$ by
Eq.~(\ref{eq:v2Fitinvmas}) in each $p_T$ bin, with $v_{2S}$
as a free parameter. Figure~\ref{v2invFit} shows $\langle
\cos[2(\varphi_{K^+K^-}-\Psi_{2})]\rangle$ [i.e. $v_{2}$ in
Equation~\ref{eq:v2Fitinvmas2}] vs $m_{inv}$ for 0.5
$<$ $p_{T}$ $<$ 1.0 GeV/$c$ in Au+Au 200~GeV collisions (0-80\%),
where the solid curve is the result of fitting
Eq.~(\ref{eq:v2Fitinvmas}). At the same time, $\langle
\sin[2(\varphi_{K^+K^-}-\Psi_{2})] \rangle$ vs $m_{inv}$
(open points) is found, as expected, to be consistent with zero
due to collisional geometry symmetry~\cite{Poskanzer:1998yz}. The
$v_{2S}$ value [i.e., $v_{2}^{obs}$ in
Eq.~(\ref{eq:v2FitFunc})] determined by the fit was corrected
for the reaction plane resolution to obtain the final $v_{2}$ for
the $\phi$ meson. The final $v_{2}$ results and related discussions will be presented  in Sec. III~E.

\begin{center}
\textbf{H.~Systematic uncertainties}
\end{center}

Major contributions to the systematic uncertainties come from variations in the procedure used for extracting the yields
from the $K^+K^-$ invariant mass distributions and from variations in the determination of tracking and particle
identification efficiencies. Different residual background functions (first-order vs. third-order polynomial
curves) were used to estimate the uncertainty of the raw yield extraction in each bin, and it was found to be of
the order of $\sim$ 4.5\%. The uncertainty due to different mixed-event normalization factors was estimated to be $\sim$ 2.1\% by
varying the normalization region in the mixed-event background distribution. The uncertainty from tracking and PID
efficiencies was estimated to be $\sim$ $8\%$, by varying the kinematic and PID cuts on the daughter tracks.

The overall systematic uncertainty was
estimated to be approximately $10\%$ for the yield ($dN/dy$), and $10\%$ for $\langle p_T \rangle$ for the Au+Au and the $d$+Au data. It includes an additional
contribution from the difference between exponential and  Levy function fittings of the transverse mass or transverse
momentum distributions. The systematic uncertainty in the overall normalization for the $p+p$ 200~GeV data was
found to be $15\%$ for $dN/dy$ and $5\%$ for $\langle p_T \rangle$, including uncertainties due to vertex
finding and trigger inefficiency for low multiplicity events.

Systematic uncertainties for the $v_{2}$ measurement
from the two different $v_{2}$ extraction methods show $p_T$ and centrality dependences, which mainly result
from the determination of S/(S+B) ratios for the invariant mass method and from the removal of residual
background in the reaction plane method, respectively. In our analysis, the point-to-point systematic errors included contributions from the following:
\begin{itemize}
\item[(i)]Difference in finding the $\phi$-meson signal via bin-by-bin counting  or Breit-Wigner function
fitting methods;

\item[(ii)]Difference due to the residual background fitting function: first- or third-order polynomial functions;

\item[(iii)]Difference in combinatorial background determination: rotation of the background (the mixed event is from the azimuthal angle rotation of all tracks from the same event) or event-mixing (the current method);

\item[(iv)]Difference in $v_{2}$ calculation: centrality-by-centrality $v_{2}$ calculation and then weighting to get the final MB $v_{2}$ or direct calculation of the $v_{2}$ through MB raw yield fitting.
\end{itemize}

\begin{center}
\textbf{ III.~RESULTS}\end{center}

\begin{center}
\textbf{A.~ Mass and width}\end{center}

\begin{figure*}[htb]
\begin{minipage}[t]{80mm}
 \includegraphics[width=80mm]{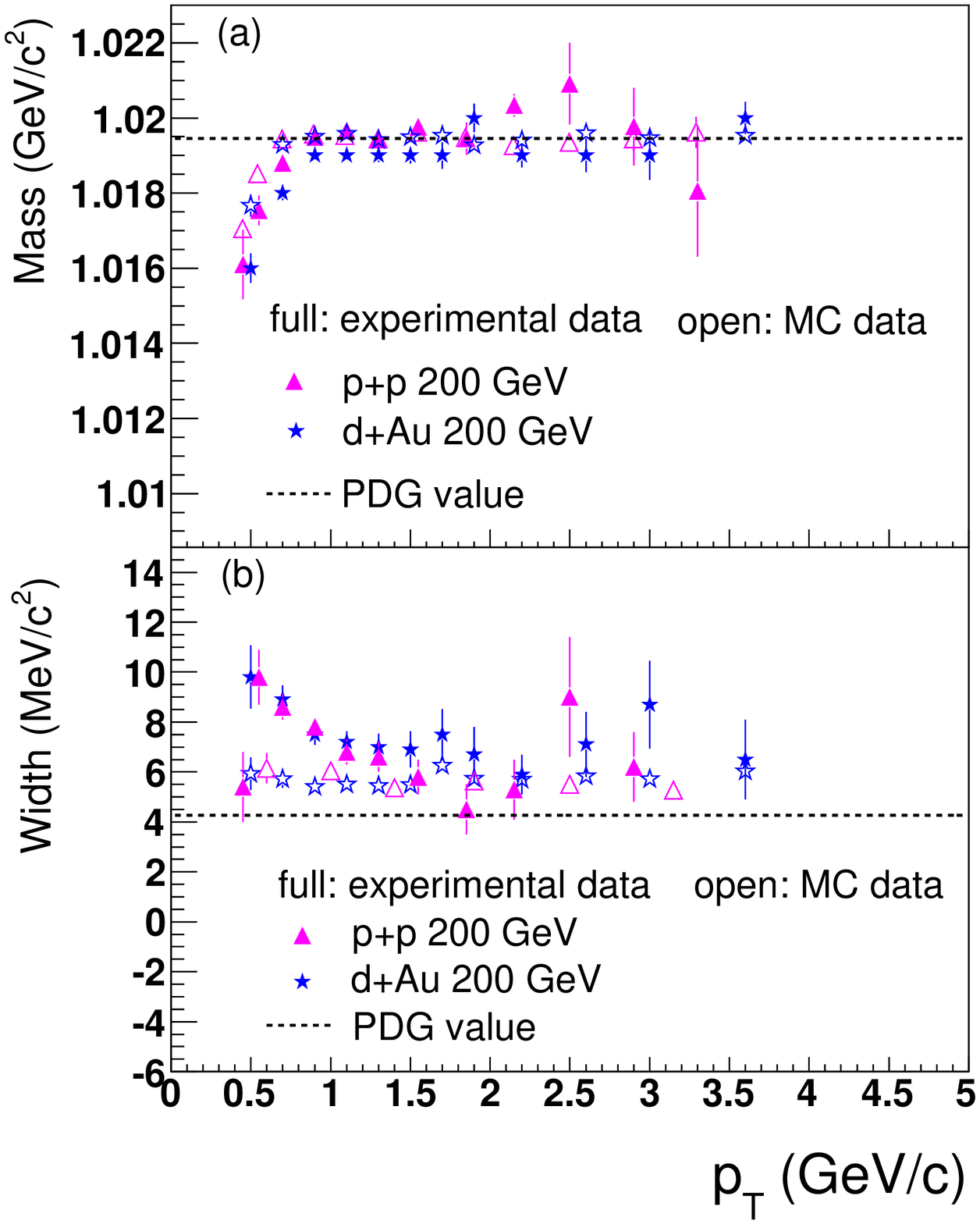}
\end{minipage}
\begin{minipage}[t]{80mm}
\includegraphics[width=80mm]{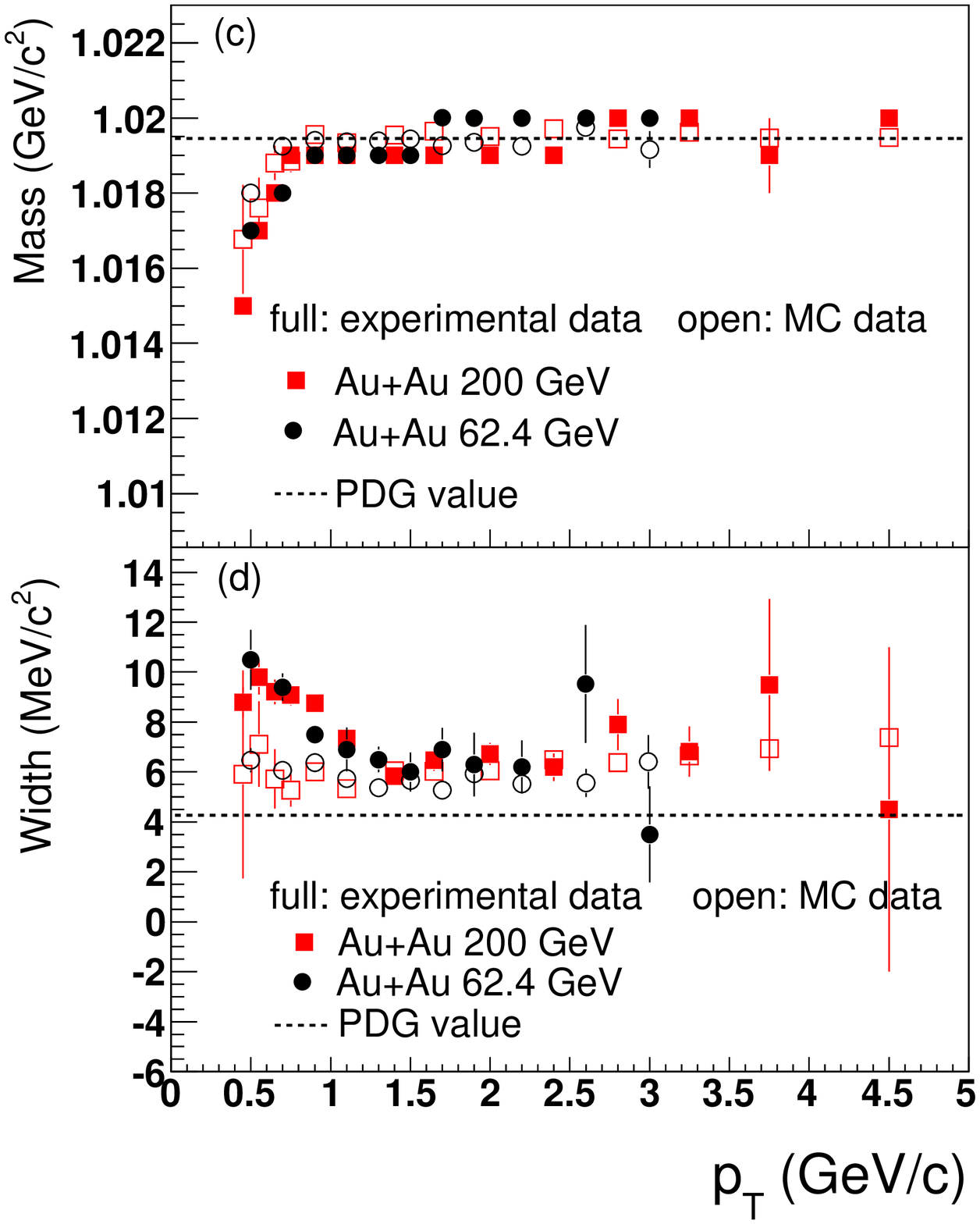}
\end{minipage}
\caption{\footnotesize(Color online) Masses and widths (FWHMs)
of $\phi$ as a function of $p_{T}$ in $p+p$ 200~GeV (NSD), $d$+Au 200~GeV (0-20\%), Au+Au 62.4~GeV (0-20\%) and Au+Au 200~GeV (0-5\%) collisions, with the corresponding PDG values.} \label{masswidth}
\end{figure*}
\begin{figure}
\includegraphics[scale=0.45]{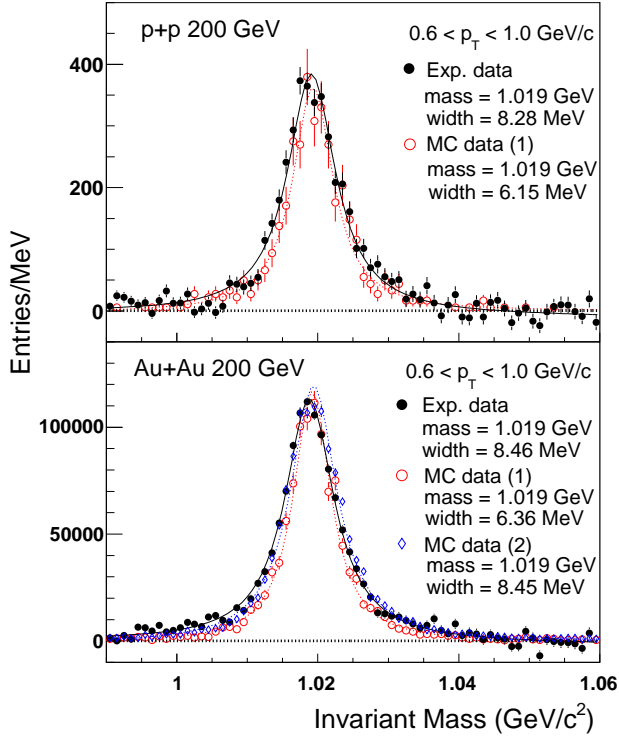}
\caption{\footnotesize(Color online) Invariant mass distributions of $\phi$ meson at  $0.6<p_{T}<1.0$ GeV/$c$ in
$p+p$ 200~GeV (NSD) and Au+Au 200~GeV (0-5\%) collisions. Solid symbols:
experimental data. Open symbols: MC simulation. Curves are the results of a Breit-Wigner function fit. Note: Two sets
of MC data are shown for Au+Au 200~GeV, and see text for details.} \label{massshapeCom}
\end{figure}

Figure~\ref{masswidth} shows the $\phi$ invariant mass peak
position and width (FWHM) as a function of $p_{T}$
for Au+Au 200~GeV (0-5\%), Au+Au 62.4~GeV (0-20\%), $d$+Au
200~GeV (0-20\%) and $p+p$ 200~GeV (NSD) collisions. In the larger
$p_T$ region ($>1$~GeV/$c$), the measured mass and width for the
$\phi$ meson are consistent with those from Monte Carlo (MC)
embedding simulations in various collision systems and at different
energies. At low $p_T$ ($<$ 1~GeV/$c$), the measured $\phi$ meson mass
is lower and the width is larger than from simulation. The drop of the $\phi$ meson mass in both real data and simulation at low $p_T$ is due to the
multiple scattering energy loss of low $p_T$ tracks in the
detector, which is not fully corrected during track
reconstruction.

Figure~\ref{massshapeCom} shows shape comparisons between
experimental and MC invariant mass distributions for the $\phi$
meson at $0.6<p_{T}<1.0$ GeV/$c$ in $p+p$ 200~GeV (NSD) and Au+Au
200~GeV (0-5\%) collisions. The real data $\phi$ invariant mass
peak (solid circles) is wider than that from standard MC data set
(1) (open circles). If the momentum resolution for low $p_T$ kaons
used in the simulations is increased by 50\%, e.g.,  kaon momentum
resolution at 350~MeV/c increases from $\sim$2\% [MC data set (1)]
to $\sim$3\% [MC data set (2)], the $\phi$ meson width from
simulation reproduces the measured width from real data as shown
by open diamonds. This decreased momentum
resolution in MC could be possible considering uncertainties in simulations 
for the amount of material between the TPC active volume and 
the primary collision vertex and residual geometry alignment issues. 
These remaining issues for the differences in mass and width of $\phi$ mesons between real data and simulations have limited our sensitivity to possible small modifications of $\phi$ meson properties in the medium produced at RHIC collisions. It should also be noted that to really trace down the possible
modification of the $\phi$ meson mass and width in heavy-ion
collisions, measurements through the dilepton decay channel are
needed. An interesting excess on the low-mass side of the $\phi$ meson
invariant mass peak was observed  by an $e^{+}+e^{-}$ channel in the low $\beta\gamma$ region ($\beta\gamma$ $<$ 1.25) for 12~GeV $p$+Cu interactions from the recent KEK experiment~\cite{Muto:2005za, sakuma:152302}. This may indicate a vector meson mass modification at normal nuclear density. $\phi$ measurements using the dilepton channel will hopefully be addressed in STAR in year 2010 with the time-of-flight detector upgrade under construction.

\begin{center}
\textbf{B.~Spectra}\end{center}

\begin{figure*}[htb]
\begin{minipage}[t]{80mm}
 \includegraphics[width=80mm]{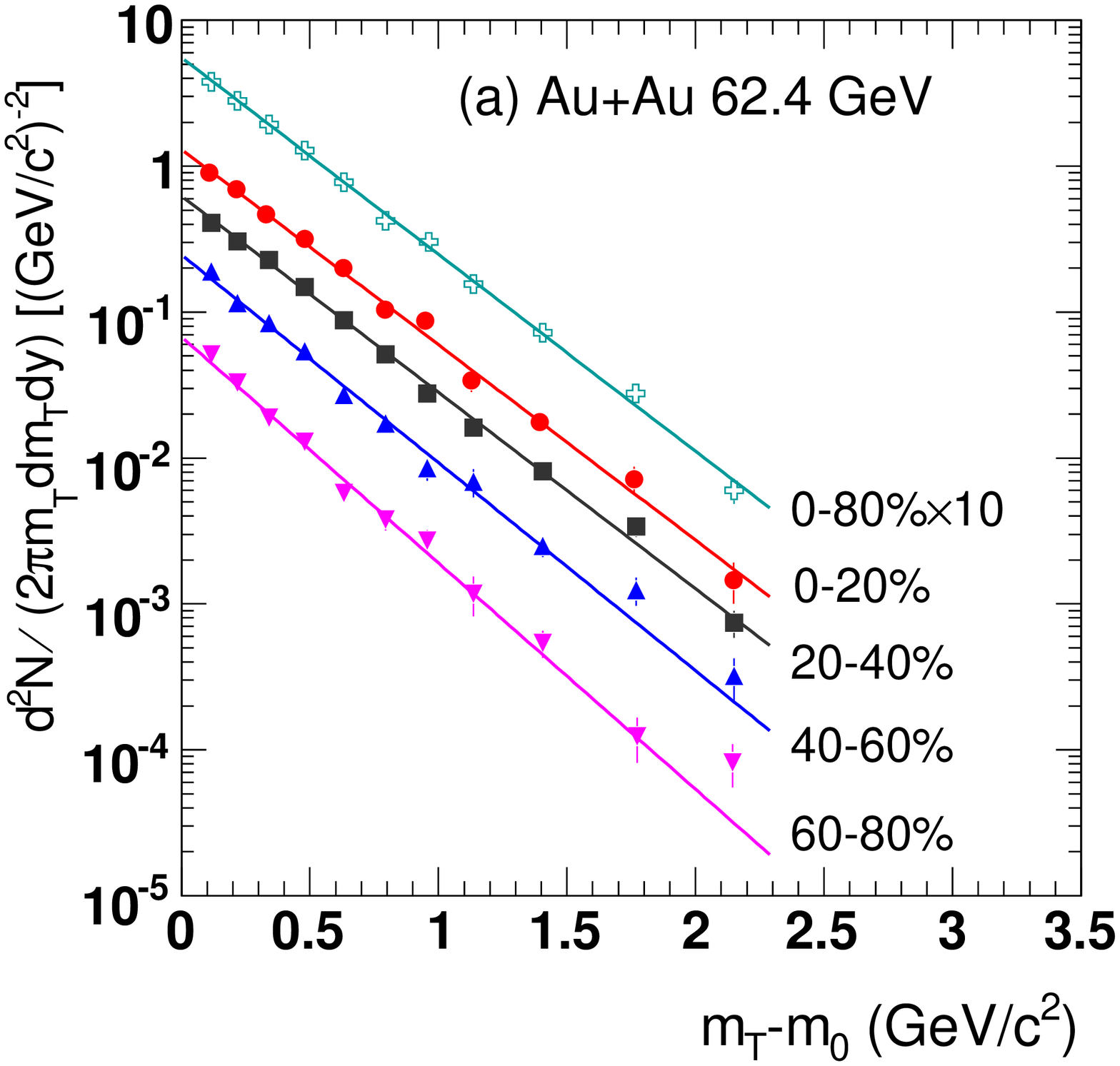}
 \includegraphics[width=80mm]{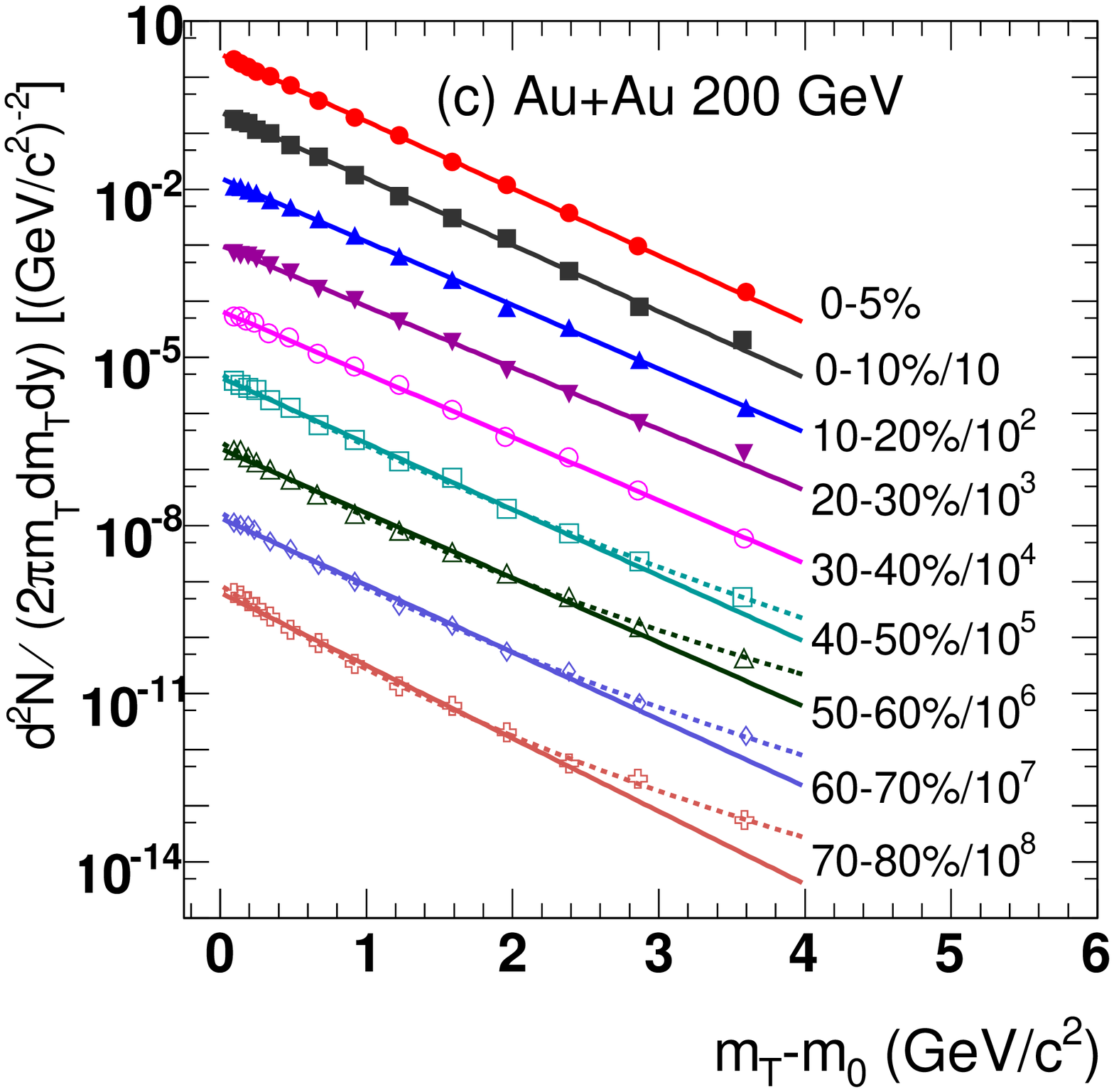}
\end{minipage}
\begin{minipage}[t]{80mm}
\includegraphics[width=80mm]{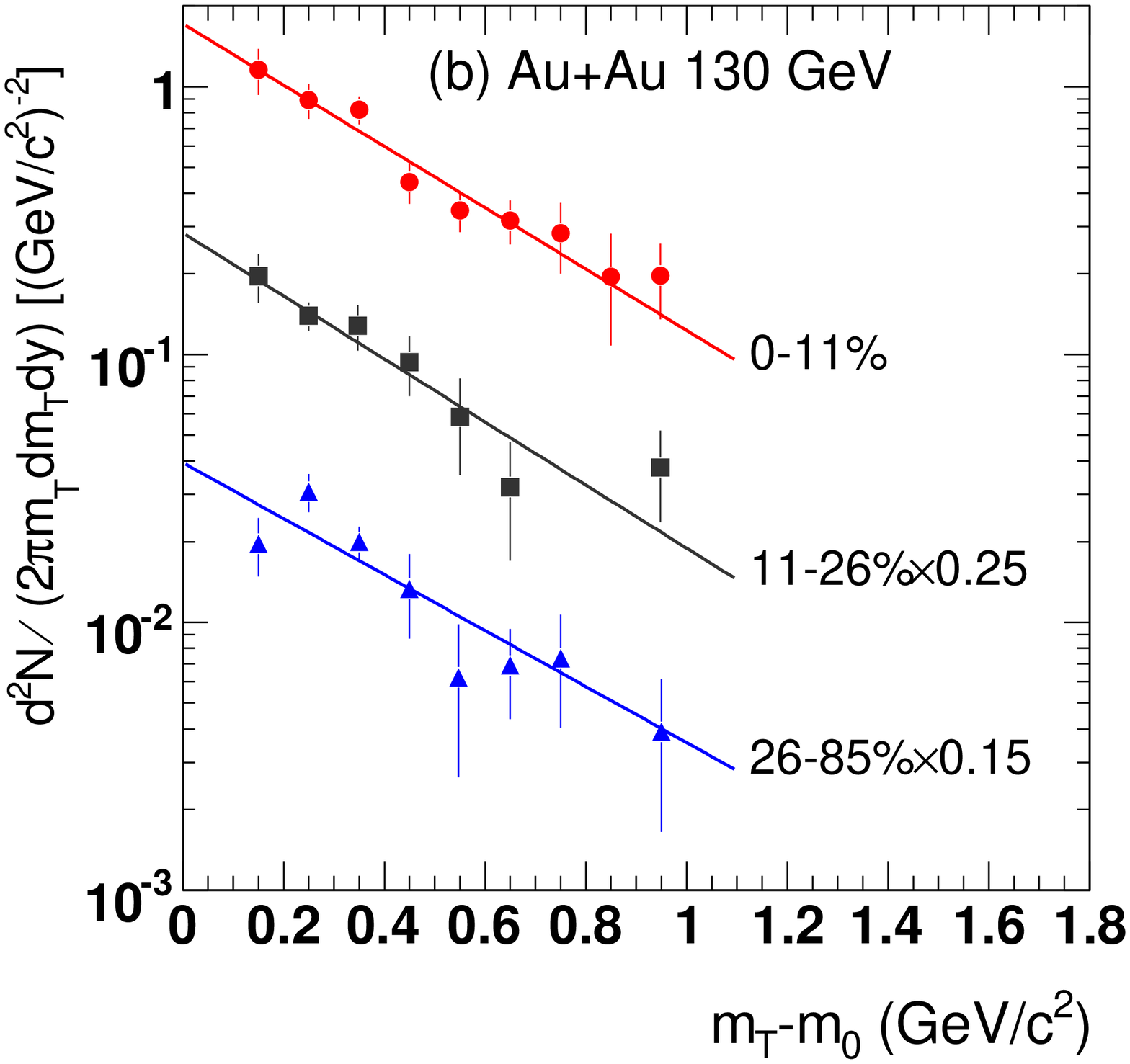}
\includegraphics[width=80mm]{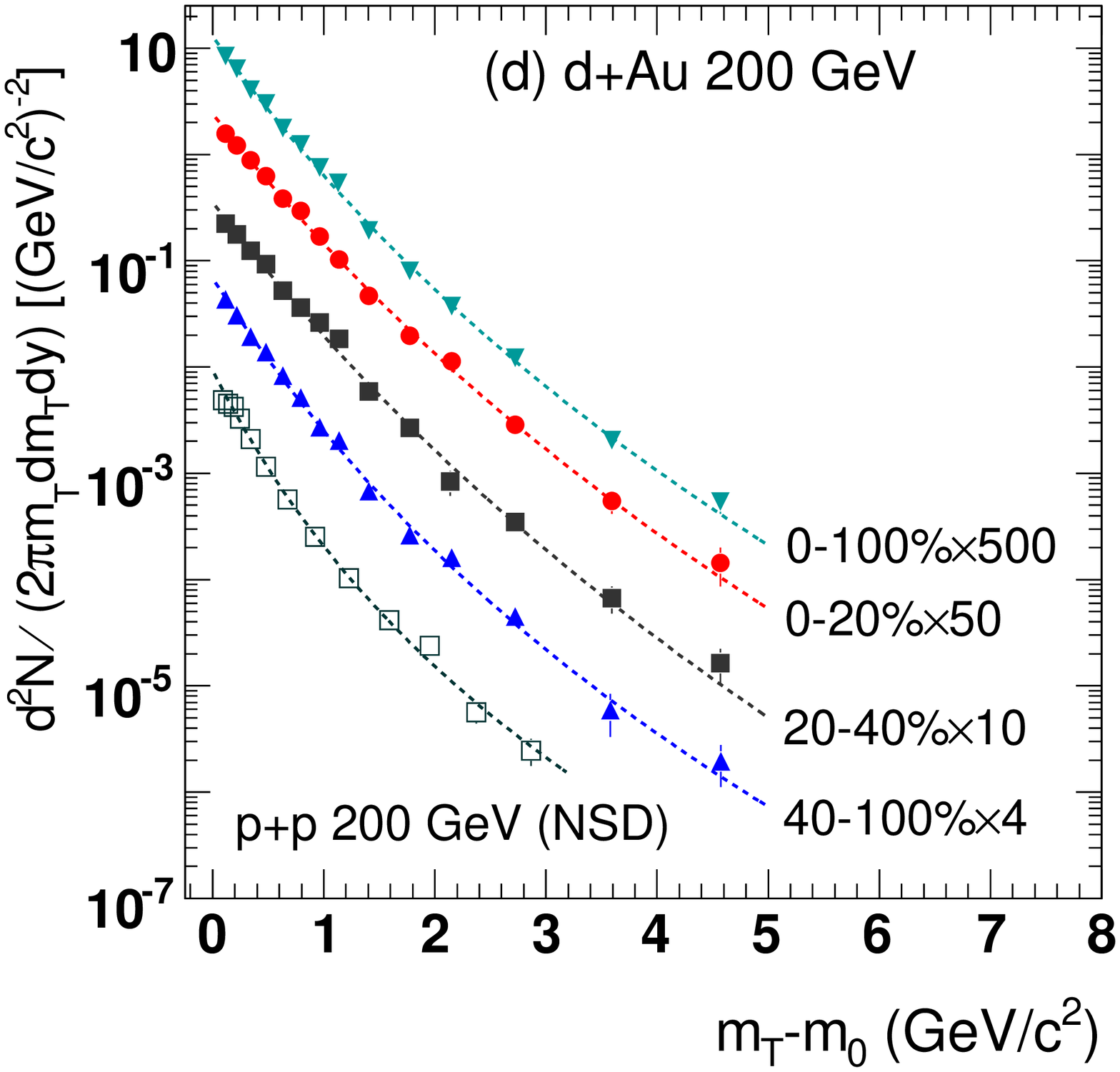}
\end{minipage}
\caption{\footnotesize(Color online) $\phi$ meson transverse mass distributions for different collision systems and different energies. For clarity, distributions for some centrality bins have been scaled by factors indicated in
the figure. Curves represent the exponential (solid) and Levy (dashed) function
fits to the distributions. Error bars are statistical only. Note that a scale factor of 1.09 is applied to the $\phi$ meson spectra for Au+Au collisions at 200 GeV in Ref.~\cite{Abelev:2007phi} to correct for the kaon identification efficiency effect missed previously.} \label{dndpt}
\end{figure*}

$\phi$ meson differential invariant yields were calculated by correcting the extracted raw yield by tracking
efficiency, detector acceptance and the decay branching ratio. Figure~\ref{dndpt} shows the $\phi$ meson
transverse mass ($m_{T}=\sqrt{p_{T}^{2}+m_{0}^{2}}$, where $m_{0}$ is the mass of $\phi$ meson) distributions from Au+Au collisions at $\sqrt{s_{_{NN}}}=$ 62.4, 130, and 200~GeV and from
$p+p$ (NSD) and $d$+Au collisions at $\sqrt{s_{_{NN}}}=$ 200~GeV. All spectra are from midrapidity, $|y|<0.5$, with $p_T$
coverage above 0.4~GeV/$c$. For clarity, distributions
for different centralities are scaled by factors indicated in the figure. Lines in the figure represent fits to
the transverse mass distributions for different centralities. The 62.4, 130 and 200~GeV Au+Au data were fitted
by the exponential function
\begin{equation}\label{exp_func}
\frac{1}{2\pi m_{T}} \frac{d^{2}N}{dm_{T} dy}=\frac{dN/dy}{2 \pi T_{exp}(m_{0}+T_{exp})} e^{-(m_{T}-m_{0})/T_{exp}},
\end{equation}
where the slope parameter $T_{exp}$ and yield $dN/dy$ are free
parameters. For $p+p$ 200~GeV and $d$+Au 200~GeV data, the
distributions were fitted by a Levy function
~\cite{Wilk:1999dr,Adams:2004ep}

\begin{eqnarray}\label{2exp_func}
    \frac{1}{2\pi m_{T}} \frac{d^{2}N}{dm_{T} dy}=\frac{dN/dy(n-1)(n-2)}{2\pi
    nT_{Levy}(nT_{Levy}+m_{0}(n-2))}\times \nonumber \\
    (1+\frac{m_{T}-m_{0}}{nT_{Levy}})^{-n},
\end{eqnarray}
where $n$, the slope parameter $T_{Levy}$, and yield $dN/dy$,
are free parameters. For the four most peripheral centrality bins
(40-50\%, 50-60\%, 60-70\% and 70-80\%) in Au+Au collisions at 200~GeV, the
distributions were better fit by a Levy rather than by an
exponential function. In fact, the exponential function
[Eq.~(\ref{exp_func})] is the limit of the Levy function
[Eq.~(\ref{2exp_func})] as $n$ approaches infinity; i.e., 
$T_{exp}$ = $T_{Levy}$($n$$\rightarrow$$\infty$).
Table~\ref{tab:fitpara} lists the extracted slope parameter $T$,
mean transverse momentum $\langle p_T \rangle$, and yield $dN/dy$ from the best fits to the spectra. Overall estimated
systematic uncertainties on these quantities are also listed.

\begin{figure}[b]
\centering
\includegraphics[width=.50\textwidth]{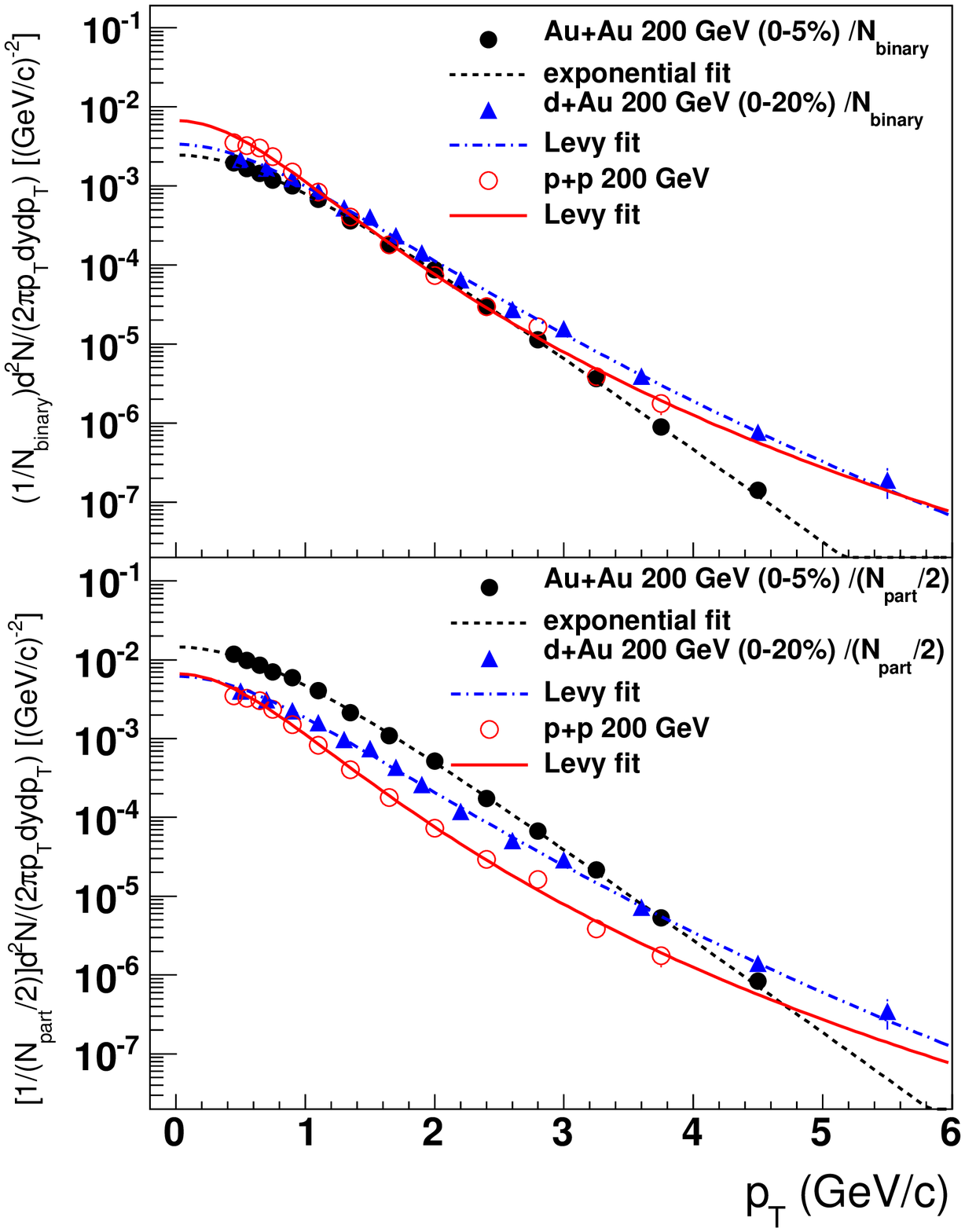}
\caption{\footnotesize(Color online) Comparison of transverse momentum spectra shape among different 200 GeV collision systems: Au+Au (0-5\%), $d$+Au (0-20\%) and $p+p$ (inelastic)). The spectra are normalized by $N_{bin}$ (top panel) and $N_{part}$/2 (bottom panel).
} \label{shapecomp}
\end{figure}

The $\phi$ meson transverse mass spectra in central Au+Au collisions can be well described by a single $m_T$-exponential
function, while the spectra in $d$+Au, $p+p$ and peripheral Au+Au are better described by a Levy function, due
to the power-law tail at intermediate and high $p_T$. Figure~\ref{shapecomp} compares the transverse momentum spectra shapes in
different 200 GeV collision systems (0-5\% Au+Au, 0-20\% $d$+Au, and
inelastic $p+p$). The spectra are normalized by the number of binary collisions
($N_{bin}$) and number of participant pairs ($N_{part}/2$). $N_{bin}$ and $N_{part}$ were determined by Glauber
model calculations~\cite{Abelev:2007}. We again point out that STAR only triggered on NSD $p+p$ events (measured
$\sigma_{NSD}=30.0\pm 3.5$~mb)~\cite{Adams:2003kv}, while the Glauber model calculations use the $p+p$ inelastic
cross section ($\sigma_{inel}=42\pm 1.0$~mb). Thus the NSD $p+p$ spectrum was normalized to the inelastic yield by a correction factor of 30/42.

A change in the shape of spectra from $p+p$, $d$+Au and peripheral Au+Au
collisions to central Au+Au collisions is clearly visible. In
comparison with the fitting result for 200~GeV $p+p$ collisions in the
high $p_T$ ($ 4.0 < p_{T} < 6.0$~GeV/$c$) region, the $N_{bin}$
normalized yield is suppressed in central Au+Au collisions at
200~GeV, while no suppression is observed for $d$+Au collisions at
200~GeV. Since particles with high transverse momentum are mostly
produced in hard scattering processes and modified by interactions
with the medium in high energy heavy-ion
collisions~\cite{Gyulassy:1990ye, Wang:1991ht, Wang:1991xy}, the
change of $\phi$ spectra from the Levy function shape in peripheral
Au+Au collisions to an exponential function shape in central Au+Au
collisions may indicate that different physics dominates the
particle production in this $p_{T}$ region. In the low $p_T$ ($p_{T}
< 1.0$~GeV/$c$) region, the $N_{part}/2$ normalized $\phi$ yield in
$d$+Au collisions scales with that in $p+p$ collisions, whereas it is
enhanced significantly in central Au+Au collisions. This indicates
that the hot environment created by central Au+Au collisions favors
the production of soft $\phi$ mesons.

Theoretical calculations have shown that particles with different
transverse momenta (or in different collision systems) are produced
by or evolve with different mechanisms, such as
hydrodynamics~\cite{Ollitrault:1992bk,Sorge:1998mk,
Huovinen:2001cy,Teaney:2000cw},
coalescence/recombination~\cite{Fries:2003vb,Fries:2003prc,
Lin:2002rw, Voloshin:2002wa, Molnar:2003ff, Greco:2003xt},
fragmentation~\cite{TSjos:1995lu,Bo:2002lu,KKP:2001NLO} and jet
quenching~\cite{Gyulassy:1990ye, Wang:1991ht, Wang:1991xy}
mechanisms. The observed change of the $\phi$ $p_{T}$ spectra shape
in our measurements is likely due to the change of these production
mechanisms in different kinematic regions and collision systems.
Further evidence of this will be discussed later, based on the
measurements of different observables.

\begin{figure}[t]
\centering
\includegraphics[width=.50\textwidth]{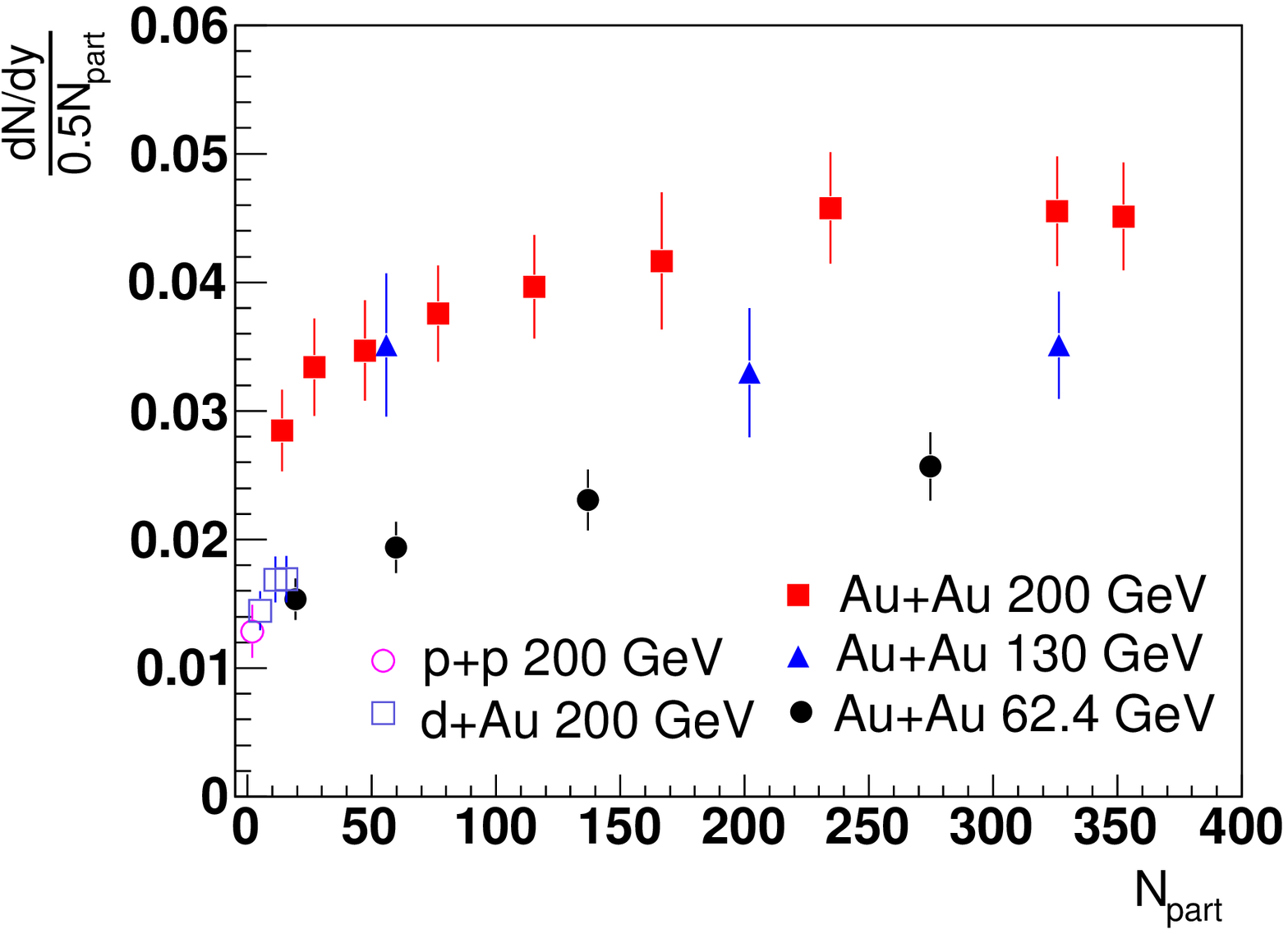}
\caption{\footnotesize(Color online) $N_{part}$ dependence of
$(dN/dy)/(0.5N_{part})$ in five different collision systems: Au+Au
62.4, 130, 200~GeV; $p+p$ 200~GeV (Inelastic); and $d$+Au
200~GeV. Statistical and systematic errors are included.}
\label{dndyanalysis2}
\end{figure}

Figure~\ref{dndyanalysis2} presents the $\phi$ meson midrapidity
yield per participant pair $(dN/dy)/(0.5N_{part})$ as a function of
$N_{part}$ (approximately proportional to the size of the collision
system). The measured midrapidity yield per participant pair
increases nonlinearly with $N_{part}$, except for the largest
centrality bins and the Au+Au 130 GeV results where there are only
three centrality bins with big error bars due to the limited
statistics. For 200~GeV
collisions, the yield increases rapidly from $p+p$ and $d$+Au to peripheral
Au+Au collisions and then saturates for midcentral Au+Au
collisions. For the same $N_{part}$, $(dN/dy)/(0.5N_{part})$
increases with the collision energy of the Au+Au collisions. This is
expected because of the increase of energy available to produce the
$\phi$ mesons. The centrality and energy dependences of
the enhancement of $\phi$ meson production can reflect the mechanism of strangeness enhancement in a dense medium formed in high energy heavy-ion collisions~\cite{phiCuCuPLB}.

\begin{figure}[t]
\includegraphics[width=.50\textwidth]{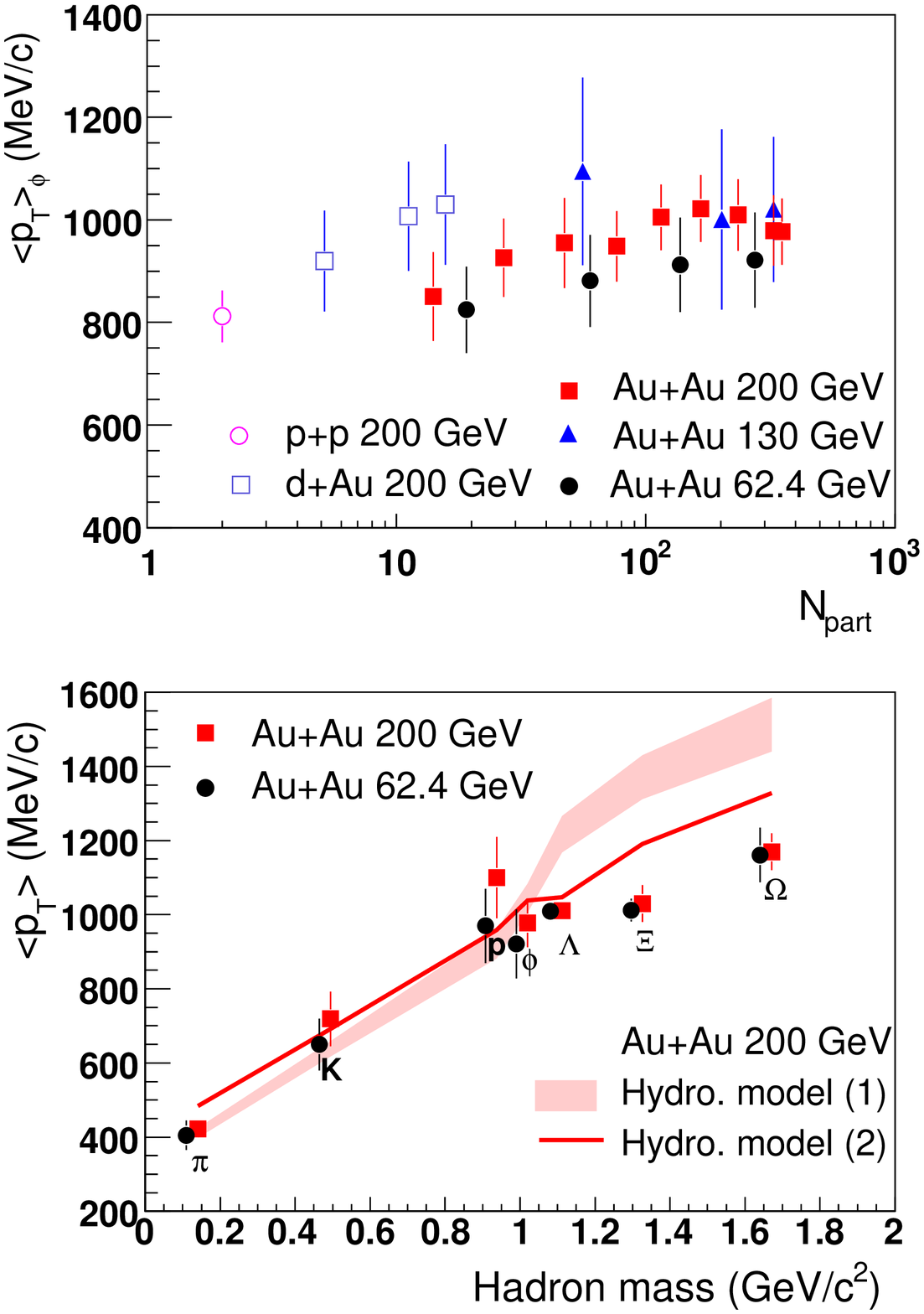}
\caption{\footnotesize(Color online) Top panel: $N_{part}$ dependence of $\langle p_{T}\rangle_{\phi}$ in different
collision systems; Bottom panel: Hadron mass dependence of $\langle p_{T}\rangle$ in central Au+Au collisions
at 62.4 and 200~GeV. The band and curve show two hydrodynamic model calculations for central Au+Au collisions at
200~GeV. Note: Hadron masses for the Au+Au 62.4~GeV data are shifted slightly in the x-axis direction for clarity,
and systematic errors are included for the $\phi$.
 }
\label{Tanalysis1}
\end{figure}

The upper panel of Fig.~\ref{Tanalysis1} shows the $N_{part}$
dependence of $\phi$ meson $\langle p_{T}\rangle$ in different
collision systems, where $\langle p_{T}\rangle$ is extracted from the
best fit to the $m_{T}$ spectra as described and shown in Table III.
The measured $\langle p_{T}\rangle$ of the $\phi$ meson shows no
significant centrality dependence within systematic errors. At the
same $N_{part}$ value, the $\phi$ meson $\langle p_{T}\rangle$
increases slightly with collision energy from 62.4 to 200~GeV.

The mean values of transverse momentum $\langle p_{T}\rangle$ as a function of  hadron mass from 62.4 and 200~GeV central Au+Au collisions
are presented in the lower panel of Fig.~\ref{Tanalysis1}. These
data are taken from Refs.~\cite{Adams:2003xp, Adams:2006yu}. The $\langle p_{T}\rangle$ of ordinary hadrons $\pi^{-}$, $K^{-}$, and
$\bar{p}$ follows a trend that is increasing with the mass of the hadron, as expected from the dynamics of these particles coming from a common radial velocity field shown as the hatched band [Hydro. model (1)] in the plot~\cite{Kolb:2003hy,Nu:2007hy}. However, heavy hyperons such as $\Xi$ and $\Omega$ show a deviation from the trend. Their values of $\langle p_{T}\rangle$ are lower than the expected ones. The observed $\langle p_{T}\rangle$ values for $\phi$ meson and $\Lambda$ are similar to those of $\Xi$ and $\Omega$.  Meanwhile, another
hydrodynamic model (2) shown by the
curve~\cite{Hama:2004hy,Qian:2007hy}, which considers possible
different chemical freeze-out temperatures for ordinary and strange
hadrons, gives a better description for strange particle $\langle
p_{T}\rangle$. This behavior can be explained if strange hadrons
have a smaller scattering cross section than ordinary hadrons in the
later hadronic stage of the collisions. These strange particles
would then decouple earlier from the system. The collective motion of the $\phi$ meson and multistrange hadrons $\Xi$ and $\Omega$ should have been developed at the early partonic stage in Au+Au collisions at RHIC. If radial flow is built
up through the evolution of the system, the particles with a
smaller hadronic cross section would have smaller radial velocity
and relatively smaller $\langle p_{T}\rangle$.

\begingroup \squeezetable
\begin{table*}
\caption{Results from fits to the transverse mass distributions of the $\phi$ meson. The fit functions used to
extract the results are also listed. All values are for midrapidity $|y|<$0.5. The first error is statistical; the second is systematic.}
\begin{ruledtabular}
\begin{tabular}{cccccccc}
   & Centrality & Fit Function & $\chi^{2}$/ndf & $T_{exp/Levy}$ (MeV) & n & $\langle p_T\rangle$ (MeV/c) & $dN/dy$ \\
  \hline
  Au+Au  & 0-20\%  & Exp. & 8.4/9 & 328$\pm$ 6 $\pm$ 22 & - & 922 $\pm$ 13 $\pm$ 61 & 3.52 $\pm$ 0.08 $\pm$ 0.45 \\
  (62.4~GeV) & 20-40\% & Exp. & 8.4/9 & 324$\pm$ 6 $\pm$ 23 & - & 913 $\pm$ 12 $\pm$ 65 & 1.59 $\pm$ 0.03 $\pm$ 0.15 \\
           & 40-60\% & Exp. & 14.5/9 & 308$\pm$ 8 $\pm$ 25 & - & 881 $\pm$ 16 $\pm$ 71 & 0.58 $\pm$ 0.01 $\pm$ 0.07\\
           & 60-80\% & Exp. & 13.3/9 & 279$\pm$ 9 $\pm$ 28 & - & 822 $\pm$ 19 $\pm$ 82 & 0.15 $\pm$ 0.004 $\pm$ 0.02\\
 \hline
  Au+Au   & 0-11\%  & Exp. & 5.3/7 & 379$\pm$ 50 $\pm$ 45 & - & 1095 $\pm$ 147 $\pm$ 131 & 5.73 $\pm$ 0.37 $\pm$ 0.57\\
  (130~GeV~\cite{Adler:2002xv}) & 11-26\% & Exp. & 3.2/5 & 369$\pm$ 73 $\pm$ 44 & - & 1001 $\pm$ 144 $\pm$ 120 & 3.33 $\pm$ 0.38 $\pm$ 0.33\\
            & 26-85\% & Exp. & 9.0/6 & 417$\pm$ 75 $\pm$ 50 & - & 1021 $\pm$ 99 $\pm$ 123 & 0.98 $\pm$ 0.12 $\pm$ 0.10\\
 \hline
  Au+Au   & 0-5\%   & Exp. & 11.0/12 & 357 $\pm$ 3 $\pm$ 23 & - & 977 $\pm$ 7 $\pm$ 64 & 7.95 $\pm$ 0.11 $\pm$ 0.73\\
  (200~GeV) & 0-10\%  & Exp. & 10.2/12 & 359 $\pm$ 5 $\pm$ 24 & - & 979 $\pm$ 20 $\pm$ 66 & 7.42 $\pm$ 0.14 $\pm$ 0.68 \\
            & 10-20\% & Exp. & 9.7/12 & 373 $\pm$ 4 $\pm$ 26 & - & 1010 $\pm$ 8 $\pm$ 69 & 5.37 $\pm$ 0.09 $\pm$ 0.50 \\
            & 20-30\% & Exp. & 26.7/12 & 387 $\pm$ 4 $\pm$ 26 & - & 1022 $\pm$ 14 $\pm$ 68 & 3.47 $\pm$ 0.06 $\pm$ 0.44\\
            & 30-40\% & Exp. & 21.1/12 & 371 $\pm$ 4 $\pm$ 24 & - & 1005 $\pm$ 8 $\pm$ 64 & 2.29 $\pm$ 0.04 $\pm$ 0.23 \\
            & 40-50\% & Levy & 17.4/11 & 315 $\pm$ 11 $\pm$ 38 & 22.7 $\pm$ 4.3 & 949 $\pm$ 13 $\pm$ 67 & 1.44 $\pm$ 0.03 $\pm$ 0.14 \\
            & 50-60\% & Levy & 6.9/11 & 290 $\pm$ 13 $\pm$ 34 & 13.8 $\pm$ 1.9 & 955 $\pm$ 14 $\pm$ 87 & 0.82 $\pm$ 0.02 $\pm$ 0.09 \\
            & 60-70\% & Levy & 7.4/11 & 291 $\pm$ 13 $\pm$ 29 & 18.6 $\pm$ 3.6 & 926 $\pm$ 15 $\pm$ 75 & 0.45 $\pm$ 0.01 $\pm$ 0.05 \\
            & 70-80\% & Levy & 5.5/11 & 243 $\pm$ 15 $\pm$ 25 & 13.0 $\pm$ 2.3 & 851 $\pm$ 19 $\pm$ 85 & 0.20 $\pm$ 0.01 $\pm$ 0.02 \\
 \hline
  $p+p$ (200~GeV, NSD~\cite{Adams:2004ux})     & 0-100\% & Levy & 10.1/10 & 202 $\pm$ 14 $\pm$ 11& 8.3 $\pm$ 1.2 & 812 $\pm$ 30 $\pm$ 41 & 0.018 $\pm$ 0.001 $\pm$ 0.003\\
 \hline
  $d$+Au    & 0-20\%   & Levy & 4.7/11 & 323 $\pm$ 20 $\pm$ 32& 15.5 $\pm$ 3.9 & 1030 $\pm$ 57 $\pm$ 103 & 0.146 $\pm$ 0.005 $\pm$ 0.014\\
  (200~GeV) & 20-40\%  & Levy & 13.6/11 & 316 $\pm$ 19 $\pm$ 32& 16.9 $\pm$ 4.7 & 1007 $\pm$ 35 $\pm$ 101 & 0.103 $\pm$ 0.003 $\pm$ 0.010\\
            & 40-100\% & Levy & 12.4/11 & 263 $\pm$ 15 $\pm$ 26& 12.2 $\pm$ 2.1 & 920 $\pm$ 35 $\pm$ 92 & 0.040 $\pm$ 0.001 $\pm$ 0.004\\
            & 0-100\% (MB) & Levy & 17.5/11 & 297 $\pm$ 11 $\pm$ 30& 13.9 $\pm$ 1.8 & 973 $\pm$ 26 $\pm$ 97 & 0.071 $\pm$ 0.001 $\pm$ 0.005\\
\end{tabular}
\end{ruledtabular}
\label{tab:fitpara}
\end{table*} \endgroup

\begin{center}
\textbf{C.~Ratios}\end{center}

\begin{figure}[htb]
\centering
\includegraphics[width=.50\textwidth]{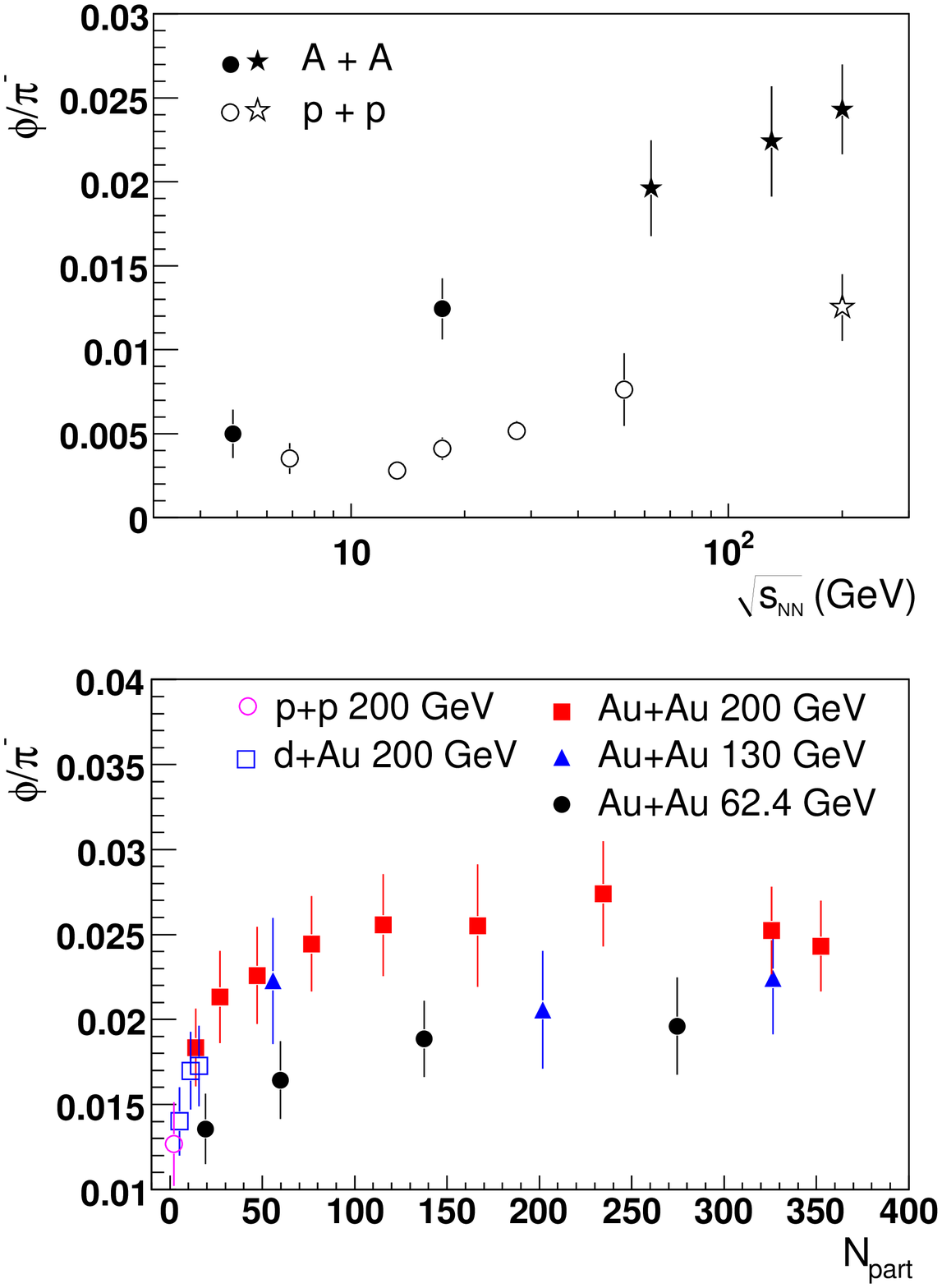}
\caption{\footnotesize(Color online) Top panel: Energy dependence
of the ratio $\phi/\pi^-$ in A+A (full points) and $p+p$ (open
points) collisions. Stars are data from the STAR experiment at RHIC.
Bottom panel: $N_{part}$ dependence of ratio $\phi/\pi^-$ in
different collision systems. Systematic errors are included for the
STAR data points.
 }
\label{phipionratio}
\end{figure}

The yield ratio $\phi/\pi^-$ as a function of the center-of-mass
energy per nucleon pair ($\sqrt{s_{_{NN}}}$) is presented in the
upper panel of Fig.~\ref{phipionratio}. The $\phi/\pi^-$ ratio
increases with energy in both A+A ~\cite{Adams:2003xp,
Adler:2002xv, Adams:2004ux, Adcox:2001mf, Back:2003rw,
Afanasiev:2000uu} and $p+p$ collisions~\cite{Blobel:1975sb,
Daum:1981tw, Aguilar-Benitez:1991yy, Drijard:1981ab}, indicating
that the yield of the $\phi$ increases faster than that of the
$\pi^-$ in A+A and  $p+p$ collisions with increasing $\sqrt{s_{_{NN}}}$. The
$\phi/\pi^-$ ratio in A+A collisions is enhanced compared to that
for $p+p$ collisions, which indicates that A+A
collisions may provide a more advantageous environment for the
production of $\phi$ mesons. In fact, an enhanced production of
$\phi$ meson in heavy-ion environment has been predicted to be a
signal of QGP formation~\cite{Shor:1984ui}. However, no clear
conclusion can be drawn from the experimental measurements since
the relative enhancement from $p+p$ to A+A collisions does not seem
to change for $\sqrt{s_{_{NN}}}>$10~GeV. The
bottom panel of Fig.~\ref{phipionratio} shows the $N_{part}$
dependence of the $\phi/\pi^-$ ratio in different collisions. The
$\phi/\pi^-$ ratio first increases with $N_{part}$, and then
seems to be saturated in the high $N_{part}$ region.

\begin{figure}[htb]
\centering
\includegraphics[width=.50\textwidth]{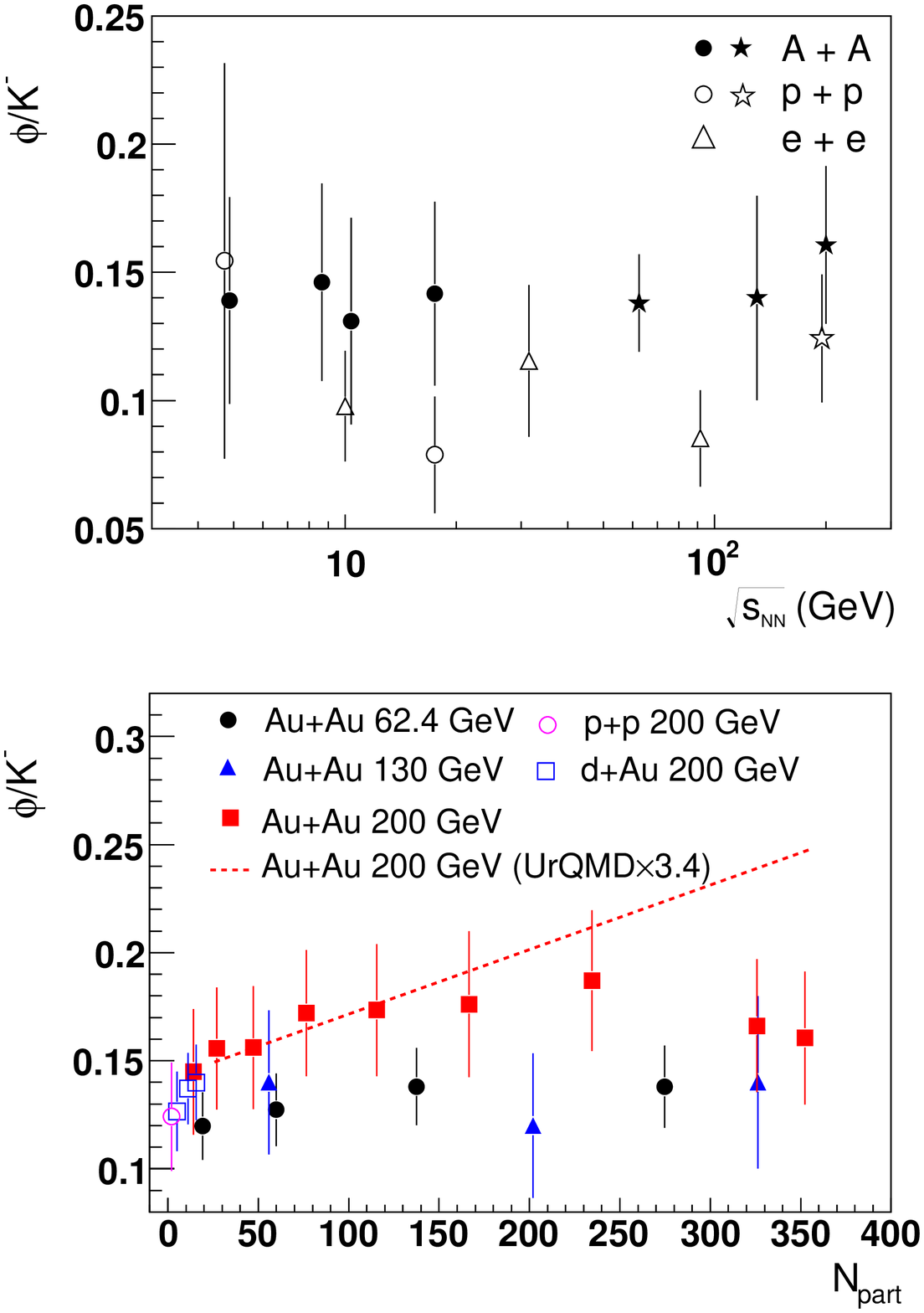}
\caption{\footnotesize(Color online) Top panel: Energy dependence
of ratio $\phi/K^-$ in A+A (full symbols) and elementary ($e+e$: open triangles and $p+p$: open circles and
star) collisions. Stars are data from STAR experiments at RHIC.
Bottom panel: $N_{part}$ dependence of ratio $\phi/K^-$ in
different collision systems. The dashed line shows results from
UrQMD model calculations. Systematic errors are included for the
STAR data points.
 }
\label{ratioanalysis}
\end{figure}

To further study whether $\phi$ meson production, or just that of
strange particles,  is enhanced in high energy A+A collisions with
respect to elementary collisions,  we have plotted the yield ratio
of $\phi/K^-$  as a function of $\sqrt{s_{_{NN}}}$ in A+A, $e+e$
and $p+p$ collisions in the top panel of Fig.~\ref{ratioanalysis}.
For these collisions, at energies above the threshold for $\phi$
production, the $\phi/K^-$ ratio is essentially independent of
collision species and energy from a few GeV up to
200~GeV~\cite{Adams:2003xp,Adler:2002xv, Adams:2004ux,
Adcox:2001mf, Back:2003rw,Afanasiev:2000uu,Blobel:1975sb,
Daum:1981tw, Aguilar-Benitez:1991yy, Drijard:1981ab}. The
lower panel of Fig.~\ref{ratioanalysis} shows that the yield
ratio $\phi/K^-$ from our analysis is also almost constant as a
function of centrality. This is remarkable considering that the
environment created by $p+p$ collisions is so drastically different
from that of Au+Au collisions that have both partonic and
hadronic interactions.

The centrality dependence of the $\phi/K^-$ ratio provides
another serious test for rescattering models based on the
assumption that kaon coalescence is the dominant mechanism for
$\phi$ production. The $K\bar{K}$ and $K$-hyperon modes are
included in these rescattering
models~\cite{Sorge:1995dp,Bleicher:1999xi}.  They predict an
increasing $\phi/K^-$ ratio vs. centrality (shown by the dashed line
in the lower panel of Fig.~\ref{ratioanalysis}), which is again
in contradiction with the approximately flat trend of our
measurements. Note also that the $\phi/K^-$ ratio from the UrQMD
model is scaled by a factor of 3.4 to match the magnitude of our
measurement for peripheral Au+Au collisions. The comparisons of
the data to predictions from these rescattering models including
$\langle p_{T}\rangle$ and $\phi/K^-$ effectively rule out kaon
coalescence as the dominant production mechanism for the $\phi$
meson.  These measurements of the $\phi/K^-$ ratio may point to a
common underlying production mechanism for $\phi$ and strange
mesons in all collision systems.

In addition, statistical models~\cite{Yen:1997rv, Wheaton:2004qb,
Braun-Munzinger:1994xr, Braun-Munzinger:1995bp,
Braun-Munzinger:1999qy} based on the assumption that the
accessible phase space is fully saturated and thermalized can
reproduce the STAR measurements of integrated hadron yield ratios
(including $\phi/\pi^-$ and $\phi/K^-$) at midrapidity in $p+p$
and Au+Au collisions at 200~GeV~\cite{Adams:2005dq}. In these
statistical models the strangeness phase-space occupancy factor
$\gamma_{s}$ approaches unity with increasing $N_{part}$,
which indicates that strangeness approaches/reaches chemical
equilibration for midcentral and central Au+Au collisions at
RHIC energies.

\begin{figure}[htb]
\centering
\includegraphics[width=0.5\textwidth]{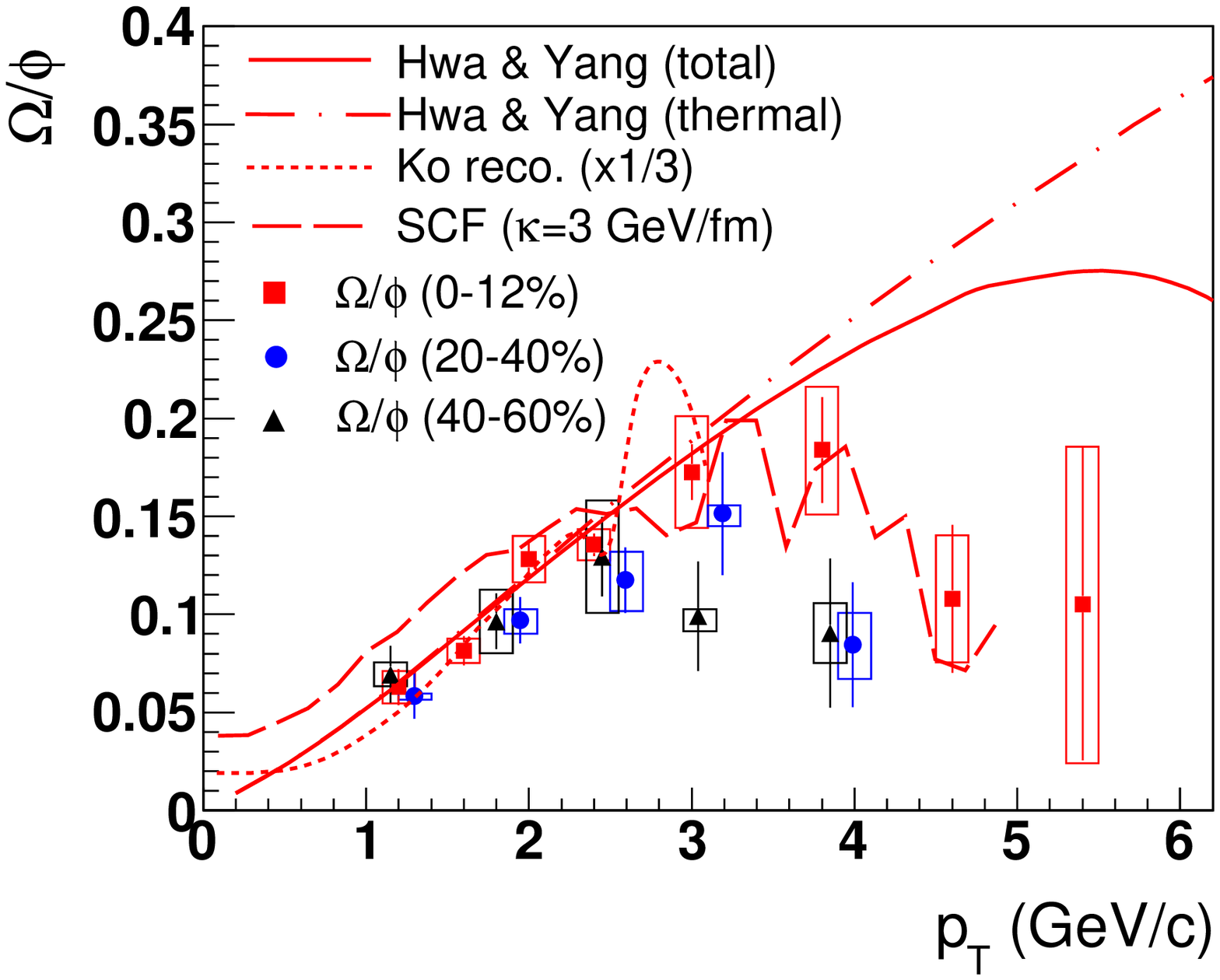}
\caption{\footnotesize(Color online) The $\Omega/\phi$ ratio
  vs.\ $p_{T}$ for three centrality bins in $\sqrt{s_{NN}}$ = 200 GeV Au+Au collisions, where the data points for 40-60\% are shifted
slightly for clarity. As shown in the legend, the lines represent the results from Hwa and Yang~\cite{Hwa:2006vb}, Ko $et~al.$~\cite{Chen:2006co} and for SCF, Refs.~\cite{Topor:2007scfprc,Armesto:2008lastcall}.}
\label{OmegaPhiRatio}
\end{figure}

Because the mechanisms of (multi)strange particle production are
predicted to be very sensitive to the early phase of nuclear
collisions, the ratio of  $\Omega/\phi$ is expected to reflect the
partonic nature of the thermal source that characterizes
QGP~\cite{Hwa:2006vb} and the effects of the strong color field
(SCF)~\cite{Topor:2007scfprc}.  In Fig.~\ref{OmegaPhiRatio}, the
ratios of $\Omega/\phi$ vs.\ $p_{T}$ are presented for different
centrality bins. The $\Omega$ ($\Omega^{-}+\overline{\Omega}^{+}$)
data points are from Ref.~\cite{Adams:2005dq} (for 0-12\%) and from
ref.~\cite{Adams:2006sa} for the other centralities. Also shown in
the figure are three curves from two recombination model
expectations for central collisions. A model  by Ko $et~al.$, based
on the dynamical recombination of quarks~\cite{Chen:2006co}, is
compared with the data and found to overpredict the ratio by a
factor of about 3 over the whole $p_{T}$ region (Note: the
bumpy shape for this model is due to the limited
statistics of their Monte Carlo simulation of the model.). Based on $\phi$ and $\Omega$ production from
coalescence of thermal $s$ quarks in the medium~\cite{Hwa:2006vb},
Hwa and Yang can describe the trend of the data up to $p_{T}\sim4$
GeV/$c$ but fail at higher $p_{T}$ (solid line). In the alternative
SCF scenario~\cite{Topor:2007scfprc,Armesto:2008lastcall}, a large string tension of
$\kappa$ = 3 GeV/fm can reproduce the data up to $p_{T}\sim4.5$ GeV/$c$
in the framework of the HIJING/$B\overline{B}$ v2.0 model, but the
effect of strong color electric fields remains an open issue and
needs to be further investigated. With decreasing centrality, the
observed $\Omega/\phi$ ratios seem to turn over at successively
lower values of $p_{T}$, possibly indicating a smaller contribution
from thermal quark coalescence in more peripheral
collisions~\cite{Hwa:2006vb}. This is also reflected in the smooth
evolution of the $p_{T}$ spectra shapes from the thermal-like
exponentials to Levy-like curves.

\begin{center}
\textbf{D.~Nuclear modification factor}\end{center}

The measurement of the nuclear modification factors $R_{cp}$ and
$R_{AB}$ provides a sensitive tool to probe the production
dynamics and hadronization process in relativistic heavy-ion
collisions~\cite{Fries:2003vb, Fries:2003prc, Lin:2002rw, Voloshin:2002wa,
Molnar:2003ff, Greco:2003xt}. $R_{cp}$, which is the ratio of yields in
central to peripheral heavy-ion collisions normalized by
$N_{bin}$, is defined as
\begin{equation}
R_{cp}(p_T) = \frac{[dN/(N_{bin} dp_T)]^{central}}{[dN/(N_{bin} dp_T)]^{peripheral}}, \label{eq:rcp}
\end{equation}
and $R_{AB}$ , which is the yield ratio of  nucleus (A) +  nucleus
(B) collisions to inelastic $p+p$ collisions normalized by
$N_{bin}$, is defined as
\begin{equation}
R_{AB}(p_T) = \frac{[dN/(N_{bin} dp_T)]^{A+B}}{[dN/ dp_T]^{p+p}}, \label{eq:raa}
\end{equation}
where $N_{bin}$ is the number of binary inelastic nucleon-nucleon
collisions determined from Glauber model
calculations~\cite{Adams:2003yh, Adams:2003kv}. It is obvious that
these two ratios will be unity if nucleus-nucleus collisions are
just simple superpositions of nucleon-nucleon collisions.
Deviation of these ratios from unity would imply contributions
from nuclear or QGP effects. It should be mentioned that when using $p+p$
collisions in the $R_{AB}$ ratio, inelastic collisions should be
used instead of the measured NSD collisions, therefore a correction factor 30/42, which was discussed in the above subsection B, was applied to correct the NSD yield to the inelastic yield in $p+p$ collisions. There is a $p_T$-dependent correction to the NSD distribution from singly diffractive (SD) events, which is
small in our measured $p_T$ range, being 1.05 at $p_T = 0.4$ GeV/$c$
and unity above 1.2 GeV/$c$ as determined from PYTHIA simulations~\cite{Adams:2003kv}.

\begin{figure}[htb]
\centering
\includegraphics[width=.50\textwidth]{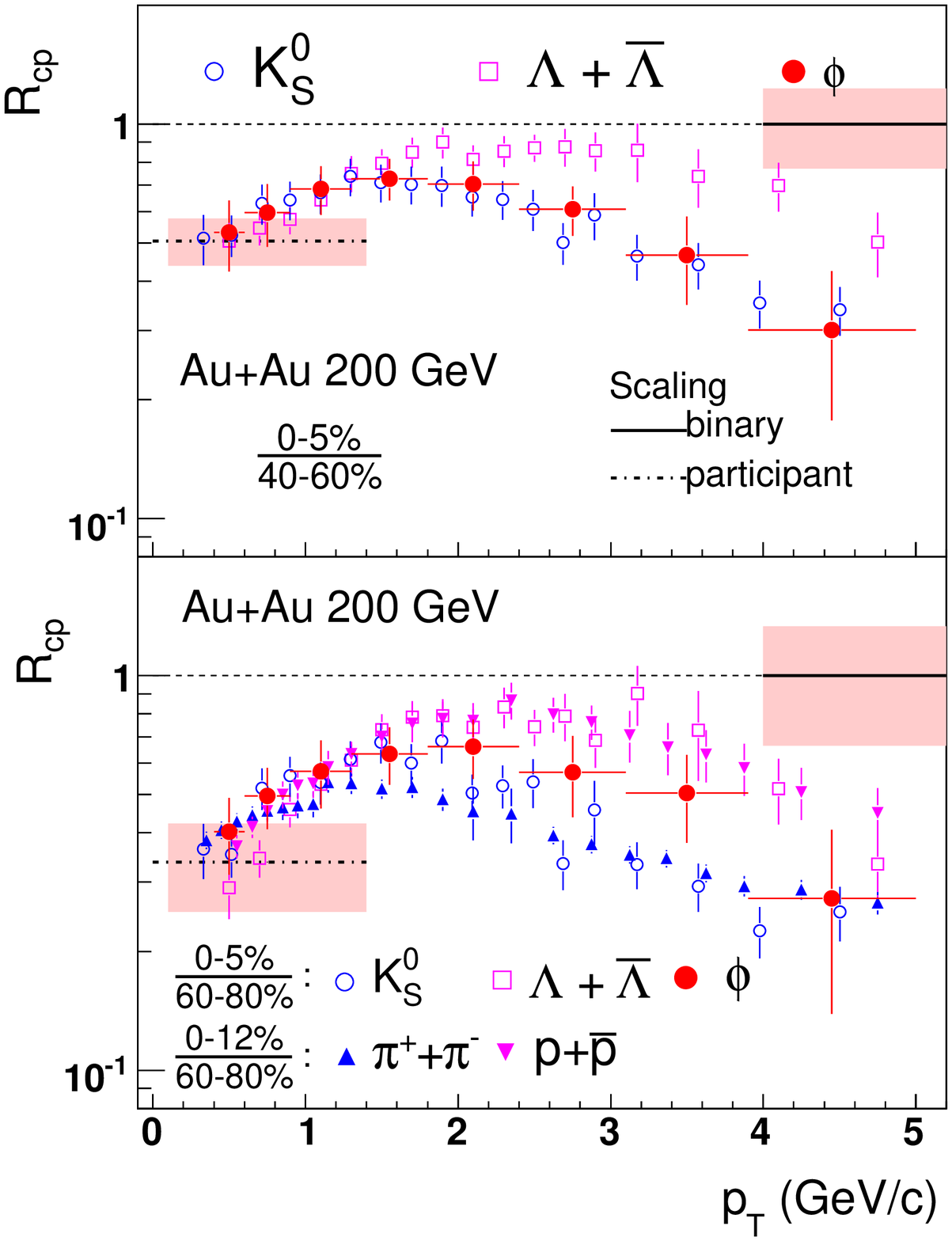}
\caption{\footnotesize(Color online) $p_{T}$ dependence of the nuclear modification factor $R_{cp}$ in Au+Au
200~GeV collisions. The top and bottom panels present $R_{cp}$ from midperipheral and most-peripheral
collisions, respectively.  See legend for symbol and line designations. The rectangular bands show the uncertainties of binary and participant scalings. Statistical and systematic errors are included.
 }
\label{rcpanalysis1}
\end{figure}

\begin{figure}[htb]
\centering
\includegraphics[width=.50\textwidth]{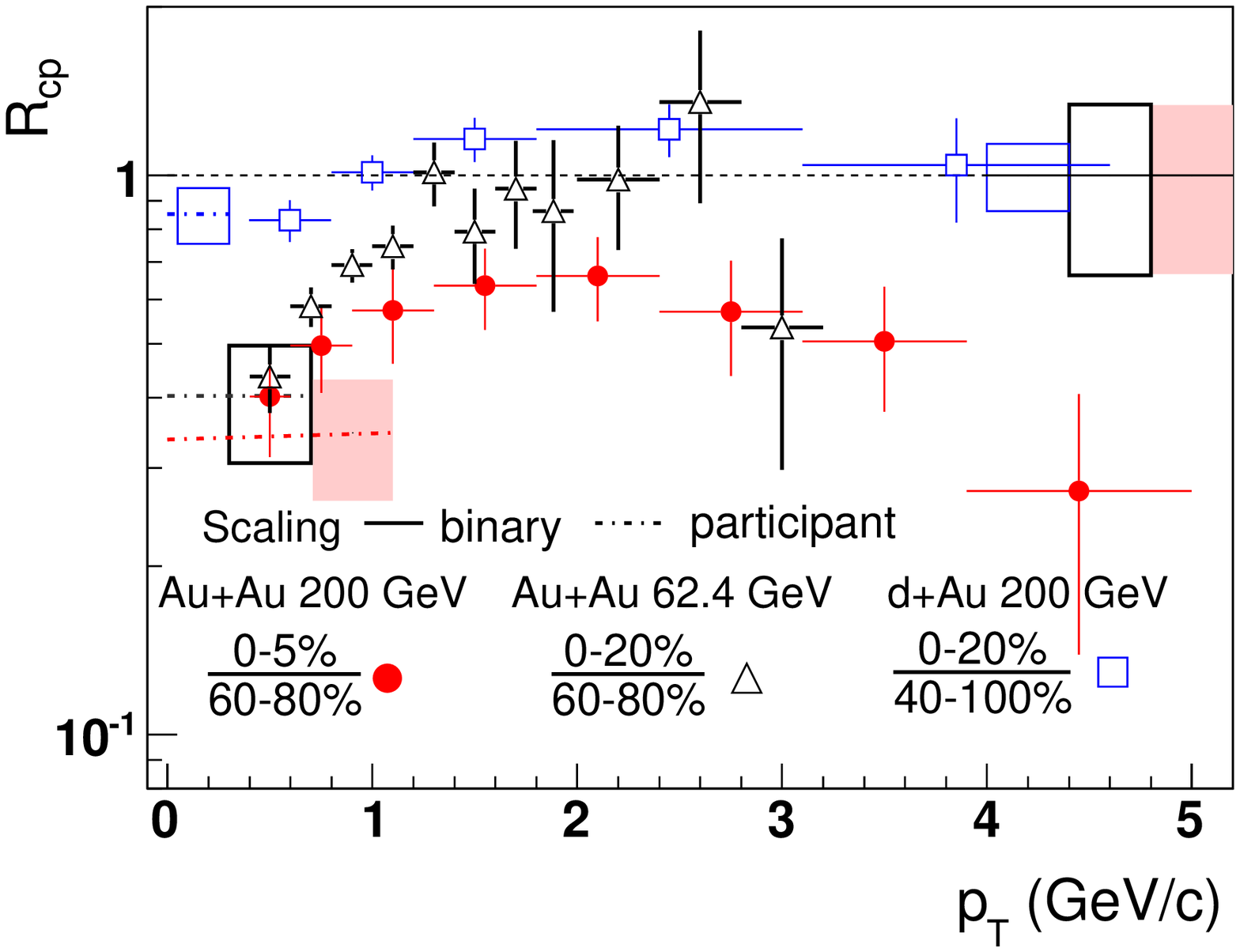}
\caption{\footnotesize(Color online) $p_{T}$ dependence of the nuclear modification factor $R_{cp}$ in Au+Au 62.4 and 200~GeV and $d$+Au 200~GeV collisions, where rectangular bands
represent the uncertainties of binary and participant scalings (see legend). Statistical and systematic errors are included.
 }
\label{rcpenergy}
\end{figure}

Figure~\ref{rcpanalysis1} presents the $p_{T}$ dependence of
$R_{cp}$ for the $\phi$ with respect to midperipheral (top panel) and most-peripheral (bottom panel) bins in Au+Au collisions at 200~GeV.
Results for the $\Lambda + \bar{\Lambda}$, the
$K^0_S$~\cite{Adams:2003am}, the $\pi^++\pi^-$ and the
$p+\bar{p}$~\cite{Abelev:2006jr} particles are also shown in the figure for
comparison. Both statistical and systematic errors are included.
Most of the systematic errors cancel in the ratios; however, the
uncertainty due to particle identification from $dE/dx$ remains as the
dominant source and varies from point to point over the range
$\sim7\%-12\%$. In the measured $p_T$ region, the $R_{cp}$ of $\phi$
meson is consistent with $N_{part}$ scaling at lowest $p_T$ (dot-dashed line), and is significantly suppressed relative to the binary collision scale
(dashed horizontal line at unity) at all $p_T$ in Au+Au
collisions at 200~GeV. When compared to the STAR measured $\Lambda +
\bar{\Lambda}$ and $K^0_S$ data~\cite{Adams:2003am}, the $R_{cp}$
for $\phi$ follows that of $K^0_S$ rather than that of the
similarly massive $\Lambda$ ($\bar{\Lambda}$), especially for the
case of 0-5\%/40-60\%. For 0-5\%/60-80\%, the $R_{cp}$ of $\phi$
sits between that for the $K^0_S$ and the $\Lambda$. This may be
attributed to the shape change of the $\phi$ spectra from
exponential at 40-60\% centrality to Levy at 60-80\% centrality
as discussed above, which may be due to the change of the $\phi$
production mechanism at intermediate $p_{T}$ in different
environments  with different degrees of strangeness equilibration.

Figure~\ref{rcpenergy} presents the $p_{T}$ dependence of $R_{cp}$ for the $\phi$ meson for Au+Au 62.4, 200~GeV
and $d$+Au 200~GeV collisions. For the three collision systems, the $R_{cp}$ factor follows $N_{part}$
scaling at low $p_T$. However, the $R_{cp}$ at intermediate $p_T$ is strongly suppressed in Au+Au collisions
at 200~GeV, is weakly suppressed in Au+Au collisions at 62.4~GeV, and shows no suppression in $d$+Au
collisions at 200~GeV.

Our measured $R_{cp}$ for the $\phi$ meson further supports the
proposed partonic coalescence/recombination scenario at intermediate
$p_{T}$~\cite{Das:1977cp, Fries:2003vb,Fries:2003prc, Hwa:2003bn},
where the centrality dependence of particle yield depends on the number of constituent quarks (NCQ) rather than
on the mass of the particle. Das and Hwa~\cite{Das:1977cp} proposed a
recombination scenario to explain the leading particle effect in $p+p$
collisions, where the fragmenting partons would recombine with sea
quarks to form hadrons. The resulting $p_T$ distribution for the
leading particle is therefore determined by the fragmenting quarks.
In heavy-ion collisions, it is possible that $q\bar{q}$ ($qqq$) come
together and form a meson (baryon) due to the high density of quarks
(antiquarks) in the collision system. The abundant nearby partons
give the recombination/coalescence mechanism a comparative
advantage over the fragmentation mechanism for particle production
in the intermediate $p_T$ region, and this successfully explains the
relative enhancement of baryons in the intermediate $p_T$ region,
such as the ratios $p/\pi$ and
$\Lambda/K_{S}^{0}$~\cite{Abelev:2006jr,Adams:2006wk,Abelev:2007ot}.
Recently, Hwa and Yang used their recombination model to
successfully describe the $\phi$ meson $p_T$ spectra, but it fails
to reproduce the $p_{T}$ dependence of $\Omega/\phi$ at higher
$p_{T}$ as shown in the previous subsection. They assumed that for the
production of $\phi$ this thermal component dominates over
others involving jet shower contributions because of the suppression
of shower $s$ quarks in central Au+Au collisions~\cite{Hwa:2006vb}.
However, the partonic medium, with an intrinsic temperature, may
modify the recombination probability and momentum distribution for
the $\phi$~\cite{Fries:2003vb,Fries:2003prc, Hwa:2003bn}.

\begin{figure}[htb]
\centering
\includegraphics[width=.50\textwidth]{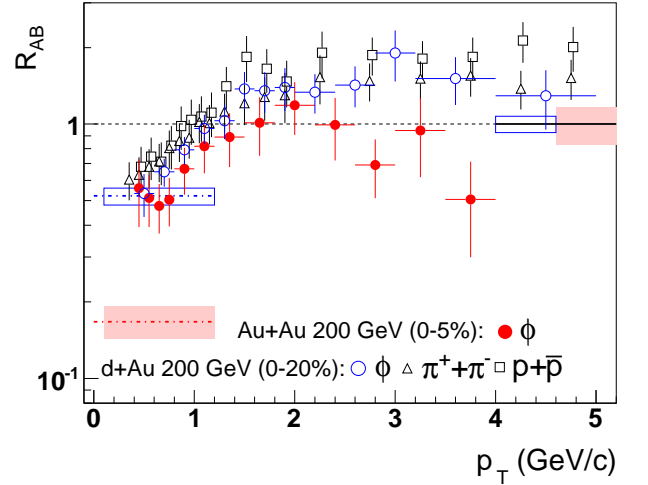}
\caption{\footnotesize(Color online) $p_{T}$ dependence of the nuclear modification factor $R_{AB}$ for $\phi$ in Au+Au
200~GeV and $d$+Au 200~GeV collisions. For comparison, data points for $\pi^++\pi^-$ in $d$+Au 200~GeV and $p+\bar{p}$ in $d$+Au 200~GeV are also shown (see legend). Rectangular bands shows the uncertainties of binary (solid line) and participant (dot-dash line) scalings. Systematic errors are included for $\phi$, $\pi^++\pi^-$ and $p+\bar{p}$.
 }
\label{raa}
\end{figure}

Figure~\ref{raa} presents the $p_T$ dependence of the nuclear
modification factor $R_{AB}$ in Au+Au and $d$+Au collisions at
200~GeV. For comparison, data points for $R_{dAu}$ of $\pi^++\pi^-$ and $p+\bar{p}$ are also shown in the figure. The $R_{dAu}$ of $\phi$ mesons reveals a similar enhancement trend as those of $\pi^++\pi^-$ and $p+\bar{p}$ at the intermediate $p_T$, which was attributed to be the Cronin effect~\cite{Cronin:1974zm,
Antreasyan:1978cw, Straub:1992xd}. The Cronin enhancement may
result either from momentum broadening due to multiple
soft~\cite{Lev:1983hh} (or semihard~\cite{Accardi:2003jh,
Papp:1999ra, Vitev:2002pf, Wang:1998ww}) scattering in the initial
state or from final state interactions suggested in the recombination
model~\cite{Hwa:2004zd}. These mechanisms lead to different particle type and/or mass dependence in the nuclear modification factors as a function of $p_T$. Our measurement of $R_{dAu}$ of $\phi$ mesons does not have the precision to differentiate particle dependence scenarios~\cite{Hwa:2005dAu,Abelev:2006dAu}.  On the other hand, the $R_{AA}$ in Au+Au 200 GeV is lower than that in $d$+Au 200 GeV and consistent with the binary collision scaling at intermediate $p_T$. These features are consistent with the scenario that features the onset of parton-medium final state interactions in Au+Au collisions. The two $R_{AB}$
observations in central $d$+Au and Au+Au collisions are consistent
with the shape comparisons in Fig.~\ref{shapecomp} (top panel).

\begin{center}
\textbf{E.~Elliptic flow}\end{center}

\begin{figure}[t]
\centering
\includegraphics[width=.50\textwidth]{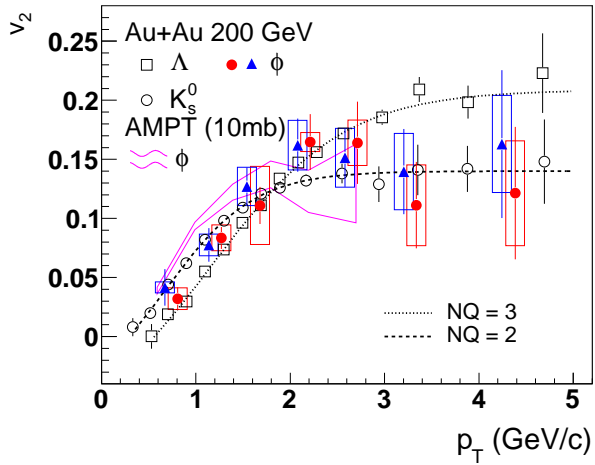}
\caption{\footnotesize(Color online) $p_{T}$ dependence of the elliptic flow $v_{2}$ of $\phi$, $\Lambda$, and $K_{S}^{0}$ in Au+Au collisions (0-80\%) at 200~GeV. Data points for $\phi$ are from the reaction plane
method (full up-triangles) and invariant mass method (full circles), where data points from the
reaction plane method are shifted slightly along the x axis for clarity. Vertical error bars represent
statistical errors, while the square bands represent systematic uncertainties. The magenta curved band
represents the $v_{2}$ of the $\phi$ meson from the AMPT model with a string melting mechanism~\cite{Chen:2006ub}. The
dash and dot curves represent parametrizations inspired by number-of-quark scaling ideas from
Ref.~\cite{Dong:2004ve} for NQ=2 and NQ=3 respectively.} \label{v2}
\end{figure}

Figure~\ref{v2} shows the elliptic flow $v_{2}$ of the $\phi$ meson as a function of $p_{T}$  in MB (0-80\%)
Au+Au collisions at $\sqrt{s_{_{NN}}}$ =200~GeV. The $v_{2}$ of the $K^0_S$ and $\Lambda$ measured by
STAR~\cite{Adams:2003am} in Au+Au 200~GeV collisions are also shown for comparison. Measurements of $v_{2}$
for the $\phi$ meson from both reaction plane and invariant mass methods are presented, and they are consistent
with each other.

The first interesting observation is that the $\phi$ meson has
significantly nonzero $v_{2}$ in the measured $p_T$ region. If the
$\phi$ meson has a small interaction cross section with the evolving hot-dense matter in A+A collisions, it would not participate in the late-stage hadronic interactions in contrast to hadrons such as $\pi$, $K$, and $p(\bar{p})$ which freeze-out later. This indicates
that the nonzero $v_{2}$ of the $\phi$ meson must have been
developed in the earlier partonic stage. In the low $p_{T}$ region
($<$2~GeV/$c$), the $v_{2}$ value of $\phi$ is between that for the
$K^0_S$ and the $\Lambda$ in Au+Au 200~GeV collisions, consistent
with the expectation of a mass ordering for $v_{2}$ in hydrodynamic
models. These observations support the hypothesis of the development
of partonic collectivity and possible thermalization in the early
stages of heavy-ion collisions at RHIC~\cite{Adams:2005dq,
Adams:2006sa}, although the underlying mechanism for the
equilibration process remains an open issue.

In the intermediate $p_{T}$ region ($\sim$2-5~GeV/$c$), the $v_{2}$
of the $\phi$ meson  is consistent with that for the $K^0_S$
rather than for the $\Lambda$. When we fit the $v_{2}(p_{T})$ of
$\phi$ mesons with the quark number scaling
function~\cite{Dong:2004ve}, the resulting fit parameter NCQ
(number of constitute quarks) = 2.3 $\pm$ 0.4. The fact that the
$v_{2}(p_{T})$ of $\phi$ is the same as that of other mesons
indicates that the heavier $s$ quarks flow as strongly as the
lighter $u$ and $d$ quarks. The AMPT model with string melting and
parton coalescence mechanisms can reproduce the experimental
results well up to 3~GeV/$c$, which favors the hadronization
scenario of coalescence/recombination of quarks~\cite{Lin:2004en,
Chen:2006ub}.

\begin{figure}[t]
\centering
\includegraphics[width=.50\textwidth]{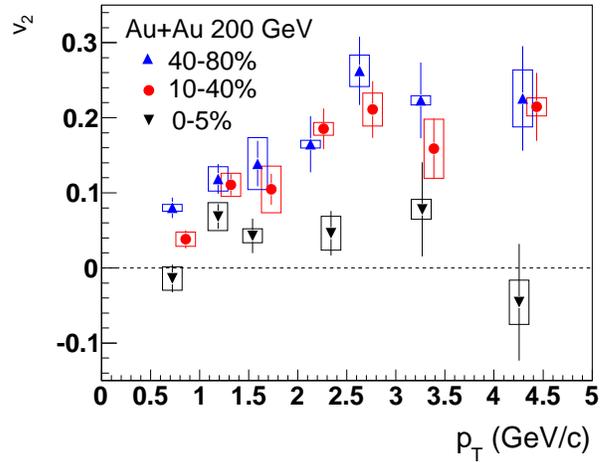}
\caption{\footnotesize(Color online) Elliptic flow $v_{2}$ as a function of $p_{T}$, $v_2(p_T)$, for the
$\phi$ meson from different centralities. The vertical error bars represent the statistical errors while the
square bands represent the systematic uncertainties. For clarity, data points of 10-40\% are shifted in the
$p_{T}$ direction slightly. } \label{v2-centrality}
\end{figure}

The $v_{2}$ of the $\phi$ meson  from other centralities are shown in
Fig.~\ref{v2-centrality}. The data are analyzed from the invariant
mass method only. As expected, $v_{2}(p_T)$ increases with
increasing eccentricity (decreasing centrality) of the initial
overlap region. This trend is also illustrated in
Table~\ref{table-v2-centrality} which presents the
$p_T$-integrated values of $\phi$-meson elliptic flow, $\langle
v_{2}\rangle$, calculated by convoluting the $v_2(p_T)$ with the
respective $p_T$ spectrum for four centrality bins. It should be
noted that the centrality dependence of the $\langle v_{2}\rangle$
of the $\phi$ meson is consistent with that of the charged
hadrons~\cite{PRC-STAR-Charged-v2}.
\begin{table}[h!]
\centering \caption{ Integrated elliptic flow $\langle
v_{2}\rangle$ for the $\phi$ meson for four centrality bins in Au+Au collisions at 200 GeV.} \label{table-v2-centrality}
\begin{ruledtabular}
\begin{tabular}{cc} Centrality (\%) & $\langle v_{2}\rangle$ (\%)  \\ \hline
40 -- 80  &  8.5 $\pm$ 1.1 (stat)  $\pm$ 0.2 (sys)  \\
10 -- 40  &  6.6 $\pm$ 0.8 (stat)  $\pm$ 0.2 (sys)  \\
0 -- 5    &  2.1 $\pm$ 1.2 (stat)  $\pm$ 0.5 (sys)  \\
0 -- 80  &  5.8 $\pm$ 0.6 (stat)  $\pm$ 0.2 (sys)  \\
\end{tabular}
\end{ruledtabular}
\end{table}

\begin{center}
\textbf{IV.~CONCLUSION}
\end{center}

In conclusion, STAR has measured $\phi$ meson production for 62.4,
130, 200~GeV Au+Au, 200~GeV $d$+Au, and NSD $p+p$ collisions at RHIC.
Details of the analysis method for $\phi$
meson are presented. The respective energy and $N_{part}$
dependence of the $\phi$ meson production, as well as the $p_{T}$
spectra for five collision systems, are reported.

The $\phi$ spectra in central Au+Au 200~GeV collisions are described
well by an exponential function. The spectra for $p+p$, $d$+Au and the
most peripheral Au+Au 200~GeV collisions are better described by a
Levy function due to the high-$p_T$ power-law tails. This change of
spectra shape from $p+p$, $d$+Au and peripheral Au+Au to central Au+Au
collisions is most likely due to a change of the dominant $\phi$
production mechanism in the different collision environments. 
The yield of $\phi$ mesons per participant pair increases
and saturates with the increase of $N_{part}$. It is found that the
$\langle p_{T}\rangle$ of $\phi$, $\Lambda$, $\Xi$, and $\Omega$ does
not follow the $\langle p_{T}\rangle$ vs. hadron mass trend
determined by the $\pi^{-}$, $K^{-}$, and $\bar{p}$. This may be
due to their small hadronic cross sections, which indicates that the
$\phi$ and strange hadrons can retain more information about the
early state of the collision system.

The $\phi/K^-$ yield ratios from $p+p$ and A+B collisions over a
broad range of collision energies above the $\phi$ threshold are
remarkably close to each other, indicating similar underlying
hadronization processes for soft strange quark pairs in these
collisions. The lack of a significant centrality dependence of the
$\phi/K^-$ yield ratio and $\langle p_T\rangle_{\phi}$ effectively
rules out kaon coalescence as a dominant production channel at RHIC.
The trend of the $\Omega/\phi$ ratio is consistent with recombination
models and  a strong color field scenario up to $p_{T}\sim4$ GeV/$c$
in central Au+Au collisions.

The measurement of the $\phi$ meson nuclear modification factor
$R_{AB}$ is consistent with the Cronin effect in $d$+Au 200~GeV
collisions, and with the energy loss mechanism in Au+Au 200~GeV
collisions. The $\phi$ meson was found to have nonzero $v_{2}$ in
the measured $p_T$ range. When comparing the $\phi$ meson nuclear
modification factor $R_{cp}$ (0-5\%/40-60\%) and elliptic flow
$v_{2}$ to those of the similar mass $\Lambda$ baryon, and to the
lighter $K^0_S$ meson, we see that the $\phi$ meson clearly
behaves more like the $K^0_S$ meson than the $\Lambda$ baryon.
Therefore, the processes relevant to $R_{cp}$ and $v_{2}$ at
intermediate $p_{T}$ are driven not by the mass of the particle,
but rather by the type of the particle, i.e. number of constituent
quarks (NCQ) scaling. The coalescence/recombination model provides
a fairly consistent picture to describe particle production in the
intermediate $p_T$ region over a broad range of collision energies
and system sizes at RHIC.

\begin{center}
\textbf{ACKNOWLEDGMENTS}
\end{center}

We thank the RHIC Operations Group and RCF at BNL, the NERSC Center 
at LBNL and the resources provided by the Open Science Grid consortium 
for their support. This work was supported in part by the Offices of NP 
and HEP within the U.S. DOE Office of Science, the U.S. NSF, the Sloan 
Foundation, the DFG Excellence Cluster EXC153 of Germany, CNRS/IN2P3, 
RA, RPL, and EMN of France, STFC and EPSRC of the United Kingdom, FAPESP 
of Brazil, the Russian Ministry of Sci. and Tech., the NNSFC, CAS, MoST, 
and MoE of China, IRP and GA of the Czech Republic, FOM of the 
Netherlands, DAE, DST, and CSIR of the Government of India, Swiss NSF, 
the Polish State Committee for Scientific Research,  and the Korea Sci. 
$\&$ Eng. Foundation.


\bibliography{Reference} 

\end{document}